\documentclass[11pt,a4paper]{article}
\usepackage[usenames,dvipsnames]{xcolor}
\usepackage[titletoc,title]{appendix}
\usepackage{amsmath}
\usepackage{esint}
\usepackage{graphicx}
\usepackage{caption}
\usepackage{subcaption}
\usepackage{indentfirst}

\input{epsf}

\def\rf#1{(\ref{eq:#1})}
\def\lab#1{\label{eq:#1}}

\def\br{\begin{eqnarray}}
\def\er{\end{eqnarray}}
\def\be{\begin{equation}}
\def\ee{\end{equation}}
\def\({\left(}
\def\){\right)}

\def\rlx{\relax\leavevmode}

\def\ve{\varepsilon}

\newcommand{\sbr}[2]{\left\lbrack\,{#1}\, ,\,{#2}\,\right\rbrack}

\def\IZ{\rlx\hbox{\sf Z\kern-.4em Z}}
\def\IR{\rlx\hbox{\rm I\kern-.18em R}}
\def\IC{\rlx\hbox{\,$\inbar\kern-.3em{\rm C}$}}
\def\one{\hbox{{1}\kern-.25em\hbox{l}}}

\begin{document}

\begin{titlepage}
\vspace*{-1cm}

\vskip 2cm

\vspace{.2in}
\begin{center}
{\large\bf A direct test of the integral Yang-Mills equations through $SU(2)$ monopoles}
\end{center}

\vspace{.5cm}

\begin{center}
C. P. Constantinidis~$^{\dagger}$, L. A. Ferreira~$^{\star}$ and G. Luchini~$^{\dagger}$

\vspace{.3 in}
\small

\par \vskip .2in \noindent
$^{\dagger}$ Departamento de F\'isica\\
 Universidade Federal do Esp\'irito Santo (UFES),\\
 CEP 29075-900, Vit\'oria-ES, Brazil

\par \vskip .2in \noindent
$^{\star}$ Instituto de F\'\i sica de S\~ao Carlos; IFSC/USP;\\
Universidade de S\~ao Paulo  \\
Caixa Postal 369, CEP 13560-970, S\~ao Carlos-SP, Brazil\\

\normalsize
\end{center}


\begin{abstract}

\noindent

We use the $SU(2)$ 't Hooft-Polyakov monopole configuration, and its BPS version, to test the   integral equations of the Yang-Mills theory. Those integral equations involve two (complex) parameters which do not appear in the differential Yang-Mills equations, and if they are considered to be  arbitrary it then implies that non-abelian gauge theories (but not abelian ones) possess an infinity of integral equations. For static monopole configurations only one of those parameters is relevant. We expand the integral Yang-Mills equation in a power series of that parameter and show that the 't Hooft-Polyakov monopole and its BPS version satisfy the integral equations obtained in first and second order of that expansion.  Our results points to the importance of  exploring the physical consequences of such an infinity of integral equations on the global properties of the Yang-Mills theory.

\end{abstract}
\end{titlepage}

\section{Introduction}
\label{sec:intro}
\setcounter{equation}{0}

The purpose of this paper is to perform a test of the integral equations of Yang-Mills theories, recently proposed in \cite{ym1,ym2}, using the $SU(2)$ 't Hooft-Polyakov monopole solution \cite{thooft,polyakov} as well as its exact analytical BPS version \cite{prasad,bogo}. The main motivation for such a test is that these integral equations involve two complex parameters that are not present in the Yang-Mills partial differential equations. If those parameters are arbitrary, it means that contrary to abelian electromagnetism, Yang-Mills theories possess in fact an infinity of integral equations. Indeed, by expanding the Yang-Mills integral equations in power series of those parameters, we check that the $SU(2)$ 't Hooft-Polyakov monopole, and its  BPS version, do satisfy the integral equations appearing in that expansion, up to second order in one of the parameters. The cancelations involved in such a check are highly non-trivial and give strong evidence on the arbitrariness of those parameters.

As shown in \cite{ym1,ym2} the integral Yang-Mills equations lead in a quite natural way to gauge invariant conserved charges. Such charges involve those two parameters in a way that if they are indeed arbitrary it would imply that in principle, the number of charges is infinite. However, due to some special properties of BPS  multi-dyon solutions \cite{manton, weinberg_book}, shown in \cite{us},  the higher charges are not really independent for such solutions, being in fact powers of the first ones (the electric and magnetic charges). The same is true for the $SU(2)$ 't Hooft-Polyakov monopole. There remains to be investigated if other non-BPS solutions also present such special properties or not, and so possess or not an infinity of charges.

In order to discuss the role of such parameters in a more concrete way let us start by the theory of electromagnetism described by  the Maxwell equations
\be
\partial_{\mu}f^{\mu\nu}= j^{\nu} \qquad\qquad \qquad  \partial_{\mu}{\tilde f}^{\mu\nu}=0
\lab{maxwelleqs}
\ee
where $f_{\mu\nu}=\partial_{\mu}a_{\nu}-\partial_{\nu}a_{\mu}$,  ${\tilde f}^{\mu\nu}= \frac{1}{2}\varepsilon_{\mu\nu\rho\lambda}f^{\rho\lambda}$, $j^{\mu}$ being the external four current, and $a_{\mu}$ the electromagnetic four vector potential.
The integral version of those equations is obtained through the abelian Stokes theorem for a rank-two antisymmetric tensor $b_{\mu\nu}$ on a space-time $3$-volume $\Omega$, as $\int_{\partial\Omega} b= \int_{\Omega} d\wedge b$, where $\partial\Omega$ is the border of $\Omega$. Taking $b_{\mu\nu}$ as a linear combination of $f^{\mu\nu}$ and its Hodge dual, and using \rf{maxwelleqs}, one gets
\be
\int_{\partial\Omega}\left[ \alpha\, f_{\mu\nu}+\beta\, {\tilde f}_{\mu\nu}\right]\,d\Sigma^{\mu\nu} =
\int_{\Omega} \beta\, {\tilde j}_{\mu\nu\rho} \,dV^{\mu\nu\rho}
\lab{maxwellinteqs}
\ee
where ${\tilde j}_{\mu\nu\rho} = \ve_{\mu\nu\rho\lambda} j^{\lambda}$ is the Hodge dual of the external current and $\alpha$ and $\beta$ are arbitrary parameters used in the liner combination.
By considering $\alpha$ and $\beta$ to be arbitrary, the integral equations \rf{maxwellinteqs} correspond to the four  usual integral equations of electromagnetic theory, which in fact preceded Maxwell differential equations. Indeed, taking $\alpha=0$ and $\Omega$ to be a purely spatial $3$-volume one gets the  Gauss law. On the hand, taking $\beta=0$ and $\Omega$ to be a solid cylinder with its height in the time direction, and its base on a spatial plane, one gets the Faraday law, and so on. The role of the parameters $\alpha$ and $\beta$ are not really important  here because \rf{maxwellinteqs} is linear in them. The situation becomes more complex in a non-abelian gauge theory.

The Yang-Mills theories were formulated \textit{\`a la} Maxwell in terms of partial differential equations, the so-called Yang-Mills equations\cite{ym_original}
\be
D_{\mu}F^{\mu\nu}= J^{\nu} \qquad\qquad \qquad  D_{\mu}{\tilde F}^{\mu\nu}=0
\lab{ymeqs}
\ee
 where $F_{\mu\nu}=\partial_{\mu}A_{\nu}-\partial_{\nu}A_{\mu}+i\,e\,\sbr{A_{\mu}}{A_{\nu}}$, with $e$ being the gauge coupling constant, ${\tilde F}^{\mu\nu}= \frac{1}{2}\varepsilon_{\mu\nu\rho\lambda}F^{\rho\lambda}$, $J^{\mu}$ being the external matter current, and $D_{\mu}=\partial_{\mu}+i\,e\,\sbr{A_{\mu}}{}$, and $A_{\mu}$ being the non-abelian gauge field taking value on the Lie algebra of the gauge group $G$.

 In order to construct the integral form of Yang-Mills equations \rf{ymeqs} one needs the non-abelian version of the Stokes theorem for a (non-abelian)  rank-two antisymmetric tensor $B_{\mu\nu}$ on a space-time $3$-volume $\Omega$. Even though  the non-abelian Stokes theorem for a one-form connection on a $2$-surface was known for some time, the same theorem for a two-form connection was constructed only more recently in \cite{afs1,afs2} using concepts on generalized loop spaces. Conceptually everything becomes more clear if one uses the two-form $B_{\mu\nu}$, defined on space-time, to construct a one-form connection on the generalized loop space. Using such generalized non-abelian Stokes theorem, the integral form of Yang-Mills equations were constructed in \cite{ym1,ym2}. The formulas involve path, surface and volume ordered integrals as follows.

 Consider a space-time $3$-volume $\Omega$, and choose a reference point $x_R$ on its border $\partial\Omega$. Scan $\Omega$ with closed $2$-surfaces based on $x_R$, labelled by a variable $\zeta$, such that $\zeta=0$ corresponds to the infinitesimal surface around $x_R$, and $\zeta=\zeta_0$ to the border $\partial\Omega$. Then scan each closed $2$-surface with loops, starting and ending at $x_R$, labelled by a variable $\tau$. Each loop is parameterized by a variable $\sigma$. The integral form of Yang-Mills equations \rf{ymeqs} is given by \cite{ym1,ym2}
 \begin{equation}
\lab{integral_equation}
V\(\partial\Omega\)\equiv P_2\;e^{ie\int_{\partial \Omega}d\tau d\sigma W^{-1}\left(\alpha F_{\mu\nu}+\beta \widetilde{F}_{\mu\nu}\right)W\frac{\partial x^\mu}{\partial \sigma}\frac{\partial x^\nu}{\partial \tau}}=P_3\;e^{\int_\Omega d\zeta d\tau V\mathcal{J}V^{-1}}\equiv U\(\Omega\),
\end{equation}
where $P_2$ and $P_3$ mean surface and volume ordered integration respectively, as explained above, and
\br
{\cal J}&=&
\int_{\sigma_i}^{\sigma_f}d\sigma\left\{ ie\beta {\widetilde J}_{\mu\nu\lambda}^W
\frac{dx^{\mu}}{d\sigma}\frac{dx^{\nu}}{d\tau}
\frac{dx^{\lambda}}{d\zeta} \right. \nonumber\\
&+& \left. e^2\int_{\sigma_i}^{\sigma}d\sigma^{\prime}
\sbr{\(\(\alpha-1\) F_{\kappa\rho}^W+\beta {\widetilde F}_{\kappa\rho}^W\)\(\sigma^{\prime}\)}
{\(\alpha F_{\mu\nu}^W+\beta {\widetilde F}_{\mu\nu}^W\)\(\sigma\)} \right.
\nonumber\\
&&\left. \times
\, \frac{d\,x^{\kappa}}{d\,\sigma^{\prime}}\frac{d\,x^{\mu}}{d\,\sigma}
\(\frac{d\,x^{\rho}\(\sigma^{\prime}\)}{d\,\tau}\frac{d\,x^{\nu}\(\sigma\)}{d\,\zeta}
-\frac{d\,x^{\rho}\(\sigma^{\prime}\)}{d\,\zeta}\frac{d\,x^{\nu}\(\sigma\)}{d\,\tau}\)\right\}
\lab{current}
\er
with $\widetilde{J}_{\mu\nu\lambda}=\ve_{\mu\nu\lambda\rho} \, J^{\rho}$, being the Hodge dual of the external matter current.  In order to simplify the formulas we have used the notation
\be
X^W\equiv W^{-1}\,X\,W
\lab{xwdef}
\ee
with  $X$ standing for the field tensor, its Hodge dual, or the dual of the matter currents. The quantity $W$ appearing above stands for the Wilson line, defined on a path parameterized by $\sigma$ through the equation
\be
\lab{w}
\frac{dW}{d\sigma}+ieA_\mu \frac{dx^\mu}{d\sigma}\;W = 0
\ee
and so
\be
W=1-i\,e\,\int_{\sigma_i}^{\sigma} d\sigma^{\prime} A_\mu\(\sigma^{\prime}\) \frac{dx^\mu}{d\sigma^{\prime}} +
\(i\,e\)^2\,\int_{\sigma_i}^{\sigma} d\sigma^{\prime} A_\mu \(\sigma^{\prime}\)\frac{dx^\mu}{d\sigma^{\prime}}\,\int_{\sigma_i}^{\sigma^{\prime}} d\sigma^{\prime\prime} A_\nu\(\sigma^{\prime\prime}\)\frac{dx^\nu}{d\sigma^{\prime\prime}} - \ldots
\ee
The quantity $V$, called the Wilson surface, is defined on a surface parameterized by $\sigma$ and $\tau$, through the equation
\be
\frac{dV}{d\tau}-V\;T(\tau) = 0,
\lab{veqdef}
\ee
with
\be
T(\tau)=ie \int_{\sigma_i}^{\sigma_f} d\sigma \;W^{-1}\left(\alpha F_{\mu\nu}+\beta \widetilde{F}_{\mu\nu}\right)W\frac{\partial x^\mu}{\partial \sigma}\frac{\partial x^\nu}{\partial \tau}.
\lab{ttaudef}
\ee
and the integration being on the closed loops used in the scanning of $\Omega$, as explained above. The initial and final values of $\sigma$, denoted $\sigma_i$ and $\sigma_f$ respectively, correspond to the initial and final points of the loop, which  in fact are the same point since the loop is always closed. Therefore the solution of \rf{veqdef} is the surface-ordered series
\be
\lab{v_sol}
V(\tau) = 1 + \int_{\tau_i}^\tau d\tau'\;T(\tau')+\int_{\tau_i}^\tau d\tau'\;\int_{\tau_i}^{\tau^\prime}\;d\tau^{\prime \prime}\; T(\tau^{\prime \prime})T(\tau')+\dots
\ee
The l.h.s. of \rf{integral_equation} is obtained by integrating \rf{veqdef} on the $2$-surface $\partial\Omega$, i.e. the border of $\Omega$. On the other hand the r.h.s. of  \rf{integral_equation} is obtained by integrating the equation
\be
\lab{u}
\frac{dU}{d\zeta}-\mathcal{K}\;U=0
\ee
 on the $3$-volume $\Omega$, and where
 \be
 \mathcal{K}=\int_{\Sigma} d\tau\, V\mathcal{J}V^{-1}
 \lab{mathcalkdef}
\ee
 with $\Sigma$ being the closed $2$-surfaces scanning $\Omega$, labelled by $\zeta$, and ${\cal J}$ given by \rf{current}.  The solution of \rf{u} is given by the volume-ordered series
\be
\lab{u_sol}
U(\zeta) = 1 + \int_{0}^\zeta d\zeta'\;\mathcal{K}(\zeta')+\int_{0}^\zeta d\zeta'\;\int_{0}^{\zeta'}\;d\zeta^{\prime \prime}\; \mathcal{K}(\zeta^{\prime} )\,\mathcal{K}(\zeta^{\prime\prime})+\dots
\ee

 Note that \rf{integral_equation} does  reduce    to \rf{maxwellinteqs} in the case that the gauge group $G$ is $U(1)$. However, for non-abelian gauge groups the dependence of both sides of \rf{integral_equation} on the parameters $\alpha$ and $\beta$ are highly non-linear. Indeed, if such parameters are arbitrary one can expand both sides of \rf{integral_equation} in a power series on them. The coefficient of each term of such series on the l.h.s. of  \rf{integral_equation} will have to equal the corresponding coefficient of the series on the r.h.s., leading to an infinity of integral equations. Consequently any solution of the Yang-Mills equations \rf{ymeqs} will have to satisfy such an infinity of integral equations. It is this test that we want to perform with the 't Hooft-Polyakov monopole, and its exact analytical BPS version \cite{weinberg_book}. We shall consider the $3$-volume $\Omega$ to be purely spatial, and consequently only the spatial components of the field tensor and its dual, i.e. $F_{ij}$ and ${\widetilde F}_{ij}$, $i,j=1,2,3$, will be present on both sides of \rf{integral_equation}. However,  ${\widetilde F}_{ij}$ is proportional to the electric field and so it vanishes for those static monopole solutions. In addition, only the component ${\tilde J}_{123}\sim J_0$ appears on the r.h.s. of  \rf{integral_equation}, and that vanishes because the solution is static and we shall work in the gauge where $A_0=0$. Remember that the only contribution for the matter current for such a solution comes from the triplet Higgs field $\phi$, and that  is of the form $J_{\mu}\sim \sbr{\phi}{D_{\mu}\phi}$. Therefore, all terms involving the parameter $\beta$ are not present on both sides of  \rf{integral_equation}, for static monopoles when $\Omega$ is purely spatial, and so it reduces to
 \begin{equation}
\lab{genbianchi}
P_2\;e^{i\,e\,\alpha\int_{\partial \Omega}d\tau d\sigma W^{-1} F_{ij}W\frac{\partial x^i}{\partial \sigma}\frac{\partial x^j}{\partial \tau}}=P_3\;e^{\alpha\(\alpha-1\)\,\int_\Omega d\zeta d\tau V{\hat {\mathcal{J}}}V^{-1}},
\end{equation}
with
\begin{equation}
{\hat {\mathcal{J}}}=\frac{e^2}{2}\left[\int_{\sigma_i}^{\sigma_f}d\sigma^\prime F_{k,l}^W(\sigma^\prime),\int_{\sigma_i}^{\sigma_f}d\sigma F_{ij}^W(\sigma)\right]\frac{\partial x^k}{\partial \sigma^\prime}\frac{\partial x^i}{\partial \sigma}\left( \frac{\partial x^l}{\partial \tau}(\sigma^\prime)\frac{\partial x^j}{\partial \zeta}(\sigma)- \frac{\partial x^l}{\partial \zeta}(\sigma^\prime)\frac{\partial x^j}{\partial \tau}(\sigma)\right)
\lab{jcalhatdef}
\end{equation}
where $i,j,k,l=1,2,3$, repeated indices are summed, and where we have denoted $F_{ij}^W\equiv W^{-1}\,F_{ij}\,W$. Note that we have explored the symmetry of ${\hat {\mathcal{J}}}$ in $\sigma$ and $\sigma^{\prime}$ to replace $\int_{\sigma_i}^{\sigma_f} d\sigma\,\int_{\sigma_i}^{\sigma} d\sigma^{\prime}\rightarrow \frac{1}{2}\,\int_{\sigma_i}^{\sigma_f} d\sigma\,\int_{\sigma_i}^{\sigma_f} d\sigma^{\prime}$.

 Equation \rf{genbianchi} is what we call {\em generalized integral Bianchi identity}. Note that one would   expect  the integral Bianchi identity to be  \rf{genbianchi} for $\alpha=1$, i.e.
  \begin{equation}
\lab{bianchi}
P_2\;e^{i\,e\,\int_{\partial \Omega}d\tau d\sigma W^{-1} F_{ij}W\frac{\partial x^i}{\partial \sigma}\frac{\partial x^j}{\partial \tau}}=\one
\end{equation}
 Indeed, that is what leads to the quantization of the magnetic charge. From a physical point of view it is intriguing that by rescaling the field tensor (magnetic field) as $F_{ij}\rightarrow \alpha\, F_{ij}$, leads to the appearance of a term like the r.h.s of \rf{genbianchi}, making the magnetic flux through $\partial\Omega$ to change drastically. However, the validity of \rf{integral_equation}, and so of
 \rf{genbianchi}, is guaranteed by the generalized non-abelian Stokes theorem for a two-form $B_{\mu\nu}$ and the  partial differential Yang-Mills equations \rf{ymeqs} as proved in \cite{afs1,afs2,ym1,ym2}. The intriguing non-linear phenomenon that we want to directly check in this paper,  is if one can expand both sides of \rf{genbianchi} in powers of $\alpha$, and if the $SU(2)$ 't Hoot-Polyakov monopole and its exact analytical BPS version, satisfy each one of the integral equations obtained through such an expansion.

 The paper is organized as follows. In section \ref{sec:expansion} we perform the expansion of the generalized integral Bianchi identity \rf{genbianchi} in powers of the parameter $\alpha$, and we show that each term of the expansion can be expressed solely in terms of the Wilson line operator.  In section \ref{sec:wilson} we calculate explicitly the Wilson line operator for the $SU(2)$ 't Hooft-Polyakov and BPS monopoles using a suitable scanning of surfaces and volumes. The result is quite simple and it is given in \rf{w2final}. In section \ref{sec:firstorder} we check the validity of the integral equation in first order of the $\alpha$-expansion, and in section \ref{sec:secondorder} we do the same for the integral equation in second order of that same expansion. We present our conclusions in section \ref{sec:conclusions} and in the appendix \ref{app:numerics} we give the results of the numerical calculations of the integrals needed to perform the check of the integral equations.

\section{The expansion of the Yang-Mills integral equations}
\label{sec:expansion}
\setcounter{equation}{0}

Assuming that  $\alpha$ and $\beta$ are indeed arbitrary we expand both sides of \rf{integral_equation} in power series in those parameters. As we have said the l.h.s. of \rf{integral_equation} is obtained by integrating \rf{veqdef}, and its  r.h.s. by integrating \rf{u}. By writing the expressions on the l.h.s and on the r.h.s of the integral equation \rf{integral_equation} in terms of \rf{v_sol} and \rf{u_sol}, and collecting the coefficients at first order in $\alpha$ and zeroth order in $\beta$, we get the integral equation at first order in $\alpha$
\br
\int_{\tau_i}^{\tau_f} d\tau\int_{\sigma_i}^{\sigma_f} d\sigma \; F_{\mu\nu}^W\frac{\partial x^\mu}{\partial \sigma}\frac{\partial x^\nu}{\partial \tau}\mid_{\zeta=\zeta_0}&=&ie
\int_{0}^{\zeta_0} d\zeta\int_{\tau_i}^{\tau_f} d\tau\int_{\sigma_i}^{\sigma_f}d\sigma \int_{\sigma_i}^{\sigma}d\sigma^{\prime}
\sbr{ F_{\kappa\rho}^W\(\sigma^{\prime}\)}
{ F_{\mu\nu}^W\(\sigma\)}
\nonumber\\
&&\times
\, \frac{d\,x^{\kappa}}{d\,\sigma^{\prime}}\frac{d\,x^{\mu}}{d\,\sigma}
\(\frac{d\,x^{\rho}\(\sigma^{\prime}\)}{d\,\tau}\frac{d\,x^{\nu}\(\sigma\)}{d\,\zeta}
-\frac{d\,x^{\rho}\(\sigma^{\prime}\)}{d\,\zeta}\frac{d\,x^{\nu}\(\sigma\)}{d\,\tau}\)
\lab{orderalpha}
\er
where $\zeta_0$ is the value of $\zeta$ corresponding to the closed surface $\partial\Omega$, in the scanning of the $3$-volume $\Omega$, which is the border of $\Omega$ (see explanation of the scanning in the paragraph above  \rf{integral_equation}). On the other hand, the integral equation  appearing in order $\beta$ and zeroth order in $\alpha$, in the expansion of  \rf{integral_equation} is given by
\br
 \int_{\tau_i}^{\tau_f} d\tau\int_{\sigma_i}^{\sigma_f} d\sigma \; \widetilde{F}_{\mu\nu}^W\frac{\partial x^\mu}{\partial \sigma}\frac{\partial x^\nu}{\partial \tau}\mid_{\zeta=\zeta_0}&=&
 \int_{0}^{\zeta_0} d\zeta\int_{\tau_i}^{\tau_f} d\tau
\int_{\sigma_i}^{\sigma_f}d\sigma\left\{ {\widetilde J}_{\mu\nu\lambda}^W
\frac{dx^{\mu}}{d\sigma}\frac{dx^{\nu}}{d\tau}
\frac{dx^{\lambda}}{d\zeta} \right. \nonumber\\
&+& \left. i\,e\,\int_{\sigma_i}^{\sigma}d\sigma^{\prime}
\sbr{ F_{\kappa\rho}^W\(\sigma^{\prime}\)}
{ {\widetilde F}_{\mu\nu}^W\(\sigma\)} \right.
\lab{orderbeta}\\
&&\left. \times
\, \frac{d\,x^{\kappa}}{d\,\sigma^{\prime}}\frac{d\,x^{\mu}}{d\,\sigma}
\(\frac{d\,x^{\rho}\(\sigma^{\prime}\)}{d\,\tau}\frac{d\,x^{\nu}\(\sigma\)}{d\,\zeta}
-\frac{d\,x^{\rho}\(\sigma^{\prime}\)}{d\,\zeta}\frac{d\,x^{\nu}\(\sigma\)}{d\,\tau}\)\right\}
\nonumber
\er
Note that in the case where the gauge group $G$ is the abelian group $U(1)$,  the equation \rf{orderalpha} corresponds to \rf{maxwellinteqs} for $\alpha=1$ and $\beta=0$. Equation \rf{orderbeta} corresponds to \rf{maxwellinteqs} for $\alpha=0$ and $\beta=1$. Note in addition that in the case where the $3$-volume $\Omega$ is purely spatial, the commutator term in \rf{orderalpha} involving the field tensors can be interpreted as a density of non-abelian magnetic charge associated to the gauge field configuration inside $\Omega$. The commutator term in \rf{orderbeta}  involving the field tensor and its Hodge dual can be interpreted as a density of non-abelian electric charge associated to the gauge field configuration inside $\Omega$. In the case where $\Omega$ has time components, those commutators will be associated to flows of non-abelian electric and magnetic charges. We have explored further those facts to obtain the integral form of the non-abelian Gauss, Faraday, etc., laws, and the physical implications of these new terms (the commutator terms) should be further explored in some other opportunity.

As one goes higher in the expansion, the integral equations become more and more complex. However, for the case we are considering in this paper, namely the static 't Hooft-Polyakov  monopole and its BPS version, and where the $3$-volume $\Omega$ is  purely spatial, there is an important simplification. As we have argued in the paragraph above
\rf{genbianchi},  only the spatial components of the field tensor (magnetic field)  appear  in the formulas, since the spatial components of its Hodge dual (electric field) vanish. As explained in section 2 of \cite{afs1}, or in the appendix of \cite{ym2}, if one performs an infinitesimal variation, $x^\mu\(\sigma\) \rightarrow x^\mu\(\sigma\)  + \delta x^\mu\(\sigma\) $, of the curve where the Wilson line $W$ \rf{w} is calculated, but keeping its end points fixed, the   infinitesimal variation of the Wilson line operator is given by
\be
W^{-1}\(\sigma_f\)\delta W\(\sigma_f\)= ie\;\int_{\sigma_i}^{\sigma_f}d\sigma\,W^{-1}\,F_{\mu\nu}\,W\,\frac{d\,x^{\mu}}{d\,\sigma}\,\delta x^{\nu}
\lab{varywilson}
\ee
The Wilson line operators $W$ appearing in the Yang-Mills integral equations \rf{integral_equation} are evaluated on the paths that scan  the closed surfaces which in their turn scan the $3$-volume $\Omega$. Thefore, as we vary the parameter $\tau$ which labels the loops, we vary the loop along a given surface, and so $\delta x^\mu = \frac{d x^\mu }{d \tau}\delta \tau$. When we vary the parameter $\zeta$ which labels the surfaces,  the loops vary perpendicular to that surface and so $\delta x^\mu = \frac{d x^\mu }{d \zeta}\delta \zeta$. Consequently,  from \rf{varywilson} we get the following two useful formulas
\begin{eqnarray}
\lab{trick}
ie\int_{\sigma_i}^{\sigma_f}d\sigma\;W^{-1}F_{\mu\nu}W\frac{d x^\mu}{d \sigma}\frac{d x^\nu}{d \tau}&=&
W^{-1}\frac{dW}{d\tau}
\nonumber\\
ie\int_{\sigma_i}^{\sigma_f}d\sigma\;W^{-1}F_{\mu\nu}W\frac{d x^\mu}{d \sigma}\frac{d x^\nu}{d \zeta}&=& W^{-1}\frac{dW}{d\zeta}.
\end{eqnarray}
As we have shown, for the static 't Hooft-Polyakov  monopole and its BPS version, and a purely spatial $3$-volume $\Omega$, the  integral Yang-Mills equation \rf{integral_equation} becomes the generalized integral Bianchi identity
\rf{genbianchi}. Therefore, from \rf{v_sol}, \rf{ttaudef} and \rf{trick}, one gets that the l.h.s. of
\rf{genbianchi} is given by
\br
V\(\partial\Omega\)\equiv P_2\;e^{i\,e\,\alpha\int_{\partial \Omega}d\tau d\sigma W^{-1} F_{ij}W\frac{\partial x^i}{\partial \sigma}\frac{\partial x^j}{\partial \tau}}&=& 1+
\alpha\;\int_{\tau_i}^{\tau_f} d\tau\;W^{-1}\frac{dW}{d\tau}(\tau)\mid_{\zeta=\zeta_0}
\lab{lhsbianchi}\\
&+&\alpha^2\;\int_{\tau_i}^{\tau_f} d\tau\;\int_{\tau_i}^{\tau}\;d\tau^{\prime }\; W^{-1}\frac{dW}{d\tau^{\prime} }(\tau^{\prime} )W^{-1}\frac{dW}{d\tau}(\tau)\mid_{\zeta=\zeta_0}+\ldots
\nonumber\\
&=& 1+ \alpha\, V_{(1)}+\alpha^2\, V_{(2)}+\ldots
\nonumber
\er
From \rf{trick} one gets that \rf{jcalhatdef} becomes
\br
{\hat {\mathcal{J}}}=-\sbr{W^{-1}\frac{dW}{d\tau}}{W^{-1}\frac{dW}{d\zeta}}.
\lab{jcalhatsimple}
\er
Therefore, from \rf{u_sol} and \rf{jcalhatsimple}, the r.h.s. of \rf{genbianchi} becomes
\br
U\(\Omega\)&\equiv&P_3\;e^{\alpha\(\alpha-1\)\,\int_\Omega d\zeta d\tau V{\hat {\mathcal{J}}}V^{-1}} =
1- \alpha\(\alpha-1\)\,\int_{0}^{\zeta_0} d\zeta\;\int_{\tau_i}^{\tau_f} d\tau\;
V\, \sbr{W^{-1}\frac{dW}{d\tau}}{W^{-1}\frac{dW}{d\zeta}}\, V^{-1}
\nonumber\\
&+& \left[\alpha\(\alpha-1\)\right]^2\,\int_{0}^{\zeta_0} d\zeta\;\int_{0}^{\zeta} d\zeta^{\prime}\;
\int_{\tau_i}^{\tau_f} d\tau\;\int_{\tau_i}^{\tau_f} d\tau^{\prime}\;
\(V\, \sbr{W^{-1}\frac{dW}{d\tau}}{W^{-1}\frac{dW}{d\zeta}}\, V^{-1}\)\(\tau\,,\,\zeta\)
\nonumber\\
&\times&
\(V\, \sbr{W^{-1}\frac{dW}{d\tau^{\prime}}}{W^{-1}\frac{dW}{d\zeta^{\prime}}}\, V^{-1}\)\(\tau^{\prime}\,,\,\zeta^{\prime}\)
+ \ldots
\lab{rhsbianchi}\\
&=& 1+ \alpha\, U_{(1)}+ \alpha^2\, U_{(2)}+\dots
\nonumber
\er
where $V$ in \rf{rhsbianchi} is evaluated with the same expansion as in \rf{v_sol} with $\beta=0$, and so an  expansion similar to  \rf{lhsbianchi}.

Therefore, by equating  \rf{lhsbianchi} to \rf{rhsbianchi} one gets that the term in first order in $\alpha$ leads to the integral equation
\br
V_{(1)}=\int_{\tau_i}^{\tau_f} d\tau W^{-1}\frac{dW}{d\tau}\mid_{\zeta=\zeta_0}=
\int_{0}^{\zeta_0} d\zeta\int_{\tau_i}^{\tau_f} d\tau\,
\sbr{W^{-1}\frac{dW}{d\tau}}{W^{-1}\frac{dW}{d\zeta}}=U_{(1)}.
\lab{firstorderbianchiint}
\er
Similarly, the term in order $\alpha^2$ gives the following integral equation
\br
V_{(2)}&=&\int_{\tau_i}^{\tau_f} d\tau\int_{\tau_i}^{\tau} d\tau^{\prime} \;
W^{-1}\frac{dW}{d\tau^{\prime}}\;W^{-1}\frac{dW}{d\tau}\mid_{\zeta=\zeta_0}=
-\int_{0}^{\zeta_0} d\zeta\int_{\tau_i}^{\tau_f} d\tau\,
\sbr{W^{-1}\frac{dW}{d\tau}}{W^{-1}\frac{dW}{d\zeta}}
\nonumber\\
&+&\int_{0}^{\zeta_0} d\zeta\int_{\tau_i}^{\tau_f} d\tau\int_{\tau_i}^{\tau} d\tau^{\prime}\;
\sbr{W^{-1}\frac{dW}{d\tau^{\prime}}}{\sbr{W^{-1}\frac{dW}{d\tau}}{W^{-1}\frac{dW}{d\zeta}}}
\lab{secondorderbianchiint}\\
&+&\int_{0}^{\zeta_0} d\zeta\int_{0}^{\zeta} d\zeta^{\prime}\int_{\tau_i}^{\tau_f} d\tau\int_{\tau_i}^{\tau_f} d\tau^{\prime}\;
\sbr{W^{-1}\frac{dW}{d\tau}}{W^{-1}\frac{dW}{d\zeta}}\;
\sbr{W^{-1}\frac{dW}{d\tau^{\prime}}}{W^{-1}\frac{dW}{d\zeta^{\prime}}}=U_{(2)}.
\nonumber
\er
We are going to verify if the $SU(2)$ 't Hooft-Polyakov monopole and its BPS version
\cite{thooft,polyakov,prasad,bogo,weinberg_book}, satisfy the integral equations \rf{firstorderbianchiint} and \rf{secondorderbianchiint}. Note that the only quantity appearing in  \rf{firstorderbianchiint} and \rf{secondorderbianchiint} is  the Wilson loop $W$. In the next section we evaluate it for those monopole solutions.

\section{The  Wilson loop for 't Hooft-Polyakov and BPS monopoles}
\label{sec:wilson}
\setcounter{equation}{0}

The spherically symmetric 't Hooft-Polyakov ansatz  \cite{thooft,polyakov} for a $SU(2)$ static magnetic monopole reads
\br
\phi&=& \frac{1}{e\,r}\,H\(\zeta\)\, {\hat r}\cdot  T
\nonumber\\
A_0&=&0
\lab{bpsmonopole}\\
A_i&=&-\frac{1}{e}\epsilon_{ijk}\frac{x_j}{r^2}(1-K(\zeta))\;T_k
\qquad\qquad \qquad
\nonumber
\er
with $r = \sqrt{x_1^2 + x_2^2 + x_3^2}$, ${\hat r}_i=x_i/r$, $\zeta = ear$,  $a$ being the vacuum expectation value of the Higgs field in the triplet representation, and $T_i$ being the generators of the $SU(2)$ Lie algebra
\be
\sbr{T_i}{T_j}=i\,\ve_{ijk}\,T_k
\lab{su2alg}
\ee
The exact analytical BPS monopole solution  corresponds to the  functions \cite{prasad,bogo}
\be
K(\zeta)=\frac{\zeta}{\sinh{\zeta}} \qquad\qquad\qquad \qquad
H(\zeta)=\zeta\,\coth\zeta-1
\lab{khbps}
\ee
For the 't Hooft-Polyakov monopole the functions $K(\zeta)$ and $H(\zeta)$ are obtained numerically, but they have  qualitatively the same behaviour as \rf{khbps}, i.e.  we have that $K(0)=1$, and then it decays monotonically (exponentially) to zero as $r\rightarrow \infty$, and $H(0)=0$, and then it grows monotonically with $r$ and for  $r\rightarrow \infty$, such grow is linear in $r$. The function $H(\zeta)$ will not be important in our calculations because the Higgs field does not appear in our integral equations  for the case of static solutions and for $\Omega$ being purely spatial (see  \rf{firstorderbianchiint} and \rf{secondorderbianchiint}). The important simplification we obtain in our calculations is due to the fact that $K(\zeta)$ is a monotonic function of $\zeta$, and so it admits an inverse function. We will then trade the parameter $\zeta$ by the function $K$, and our calculations will not depend upon the explicit form of the function $K(\zeta)$.

We have chosen to evaluate both sides of the integral equations  \rf{firstorderbianchiint} and \rf{secondorderbianchiint} on a purely spatial $3$-volume $\Omega$ which is a ball centered at the origin of the Cartesian coordinate system $x_i$, $i=1,2,3$, used in the ansatz \rf{bpsmonopole}. We then scan that volume $\Omega$ with closed surfaces which are spheres also centered at the origin of the Cartesian coordinate system, with radii varying from zero to the radius of $\Omega$. However, since the reference point $x_R$ have to lie on the border $\partial \Omega$ of $\Omega$, and since the surfaces scanning it have to be based at $x_R$, we shall attach to the ball $\Omega$ a infinitesimally thin cylinder  lying on the negative $x_1$-axis, and locate the reference point $x_R$ at $\(x_1\,,\,x_2\,,\,x_3\)=\(-\infty\,,\,0\,,\,0\)$. The cylinder has a radius $\ve$, which will be taken to zero at the end of the calculations. The surfaces scanning $\Omega$ will have the form depicted in Figure \ref{atwood2}, i.e. an infinitesimally thin cylinder on the negative $x_1$-axis and a sphere centered at the  origin of the Cartesian coordinate system. With the attachment of the thin cylinder we can keep the surfaces based at $x_R$, and centered at the origin.  In addition, $x_R$ being at infinity allows us to have the volume $\Omega$ with any radius. We shall label the surfaces scanning $\Omega$ with the parameter $\zeta$, which is the same as the one appearing in the ansatz \rf{bpsmonopole}. Then $\zeta=0$ corresponds to the surface made of the thin cylinder and a sphere of radius zero attached to it, and $\zeta=\zeta_0$ corresponds to the border $\partial\Omega$, made of the thin cylinder attached to a sphere of radius $\zeta_0$, the same as the radius of $\Omega$, centered at the origin. The loops will be labelled by a parameter $\tau$, they start and end at the reference point $x_R$, and there will be three types of loops, as follows:
\begin{enumerate}
 \item {\bf Loops of type (I), scanning the thin cylinder,  as depicted in Figure \ref{atwood1}}. For such loops the parameter $\tau$ varies from $-\infty$ to $-\frac{\pi}{2}$, with $\tau=-\infty$ corresponding to the infinitesimal loop around $x_R$, and $\tau=-\frac{\pi}{2}$ corresponding to a straight line from $x_R$ to the border of the sphere, then encircling the joint of the cylinder with the sphere, and coming back to $x_R$ through the same straight line. The three parts of such loops will be denoted (I.1), the first straight line, (I.2), the circle and (I.3) the second straight line. We parameterize the loops with $\sigma$, such that the points on the loops have the following coordinates:
 $$
\begin{array}{lllll}
\textrm{(I.1)}& x_1 = \tau + \sigma -\zeta + \frac{\pi}{2} &\qquad x_2 = 0 & \qquad x_3 = -\varepsilon & \qquad (-\infty \leq \sigma \leq 0) \\
\textrm{(I.2)} & x_1 = \tau - \zeta + \frac{\pi}{2} & \qquad x_2 = \varepsilon \sin\sigma & \qquad x_3 = -\varepsilon\cos\sigma & \qquad (0\leq \sigma \leq 2\pi)\\
\textrm{(I.3)} & x_1 = \tau +2\pi - \sigma -\zeta + \frac{\pi}{2} &\qquad x_2 = 0 & \qquad x_3 = -\varepsilon & \qquad (2\pi \leq \sigma \leq \infty)\\
\end{array}
$$
with fixed $\zeta$ and $-\infty \leq \tau \leq -\frac{\pi}{2}$.

\begin{figure}[htbp]
\begin{center}
\includegraphics[width=0.7\textwidth]{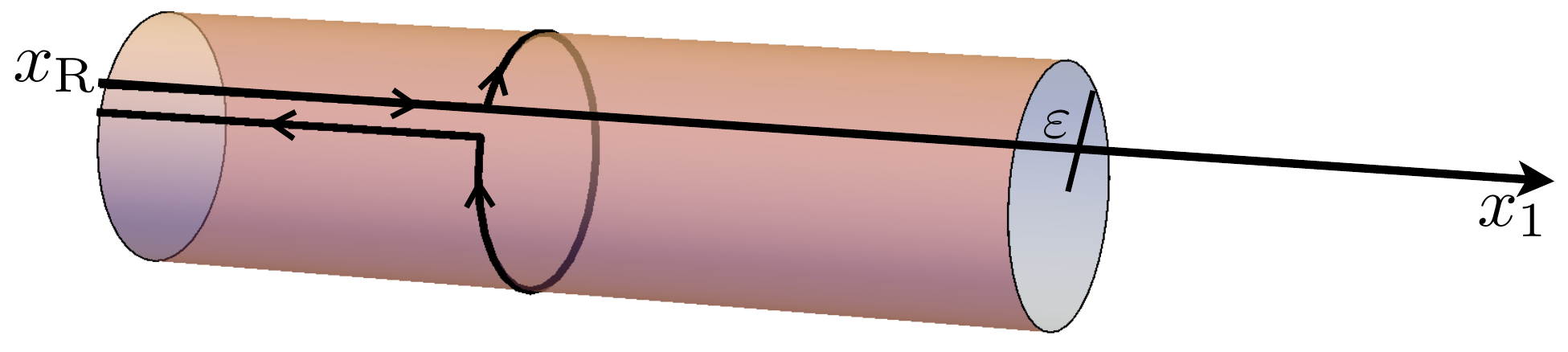}
\caption{Scanning of type (I). The gap between the straight lines is only a visual resource. }
\label{atwood1}
\end{center}
\end{figure}

 \item {\bf Loops of type (II), scanning the thin sphere,  as depicted in Figure \ref{atwood2}}. For such loops the parameter $\tau$ varies from $-\frac{\pi}{2}$ to $\frac{\pi}{2}$. A loop of this type is made of  a straight line from $x_R$ to the border of the sphere, then making a circle on the surface of the sphere, starting and ending at the junction of the cylinder with the sphere, and lying on a plane perpendicular to the plane $x_1\,x_3$,  that makes an angle $\tau$ with the plane $x_1\,x_2$. Finally, it returns to $x_R$ through the same straight line. Again, the three parts of such loops will be denoted as (II.1) for the first straight line, (II.2) for the circle and (II.3) labels the second straight line. We parameterize the loops with $\sigma$, such that the points on the loops have the following coordinates:
 $$
\begin{array}{lllll}
\textrm{(II.1)}& x_1 = \sigma - \zeta & x_2 = 0 &  x_3 = -\varepsilon &  (-\infty \leq \sigma \leq 0) \\
\textrm{(II.2)} & x_1 = \zeta\left(\cos^2\tau (1-\cos\sigma)-1\right) &  & &\\
& x_2 = \zeta \cos\tau \sin\sigma &  && (0\leq \sigma \leq 2\pi)\\
& x_3 = \zeta\cos\tau\sin\tau(1-\cos\sigma) &  &&\\
\textrm{(II.3)} & x_1 = -\sigma + 2\pi -\zeta & x_2 = 0 &  x_3 = -\varepsilon &  (2\pi \leq \sigma \leq \infty).\\
\end{array}
$$
with fixed $\zeta$ and where in $(II.2)$ the parameter $\tau$ varies from $-\frac{\pi}{2}$ to $\frac{\pi}{2}$.

\begin{figure}[htbp]
\begin{center}
\includegraphics[width=0.9\textwidth]{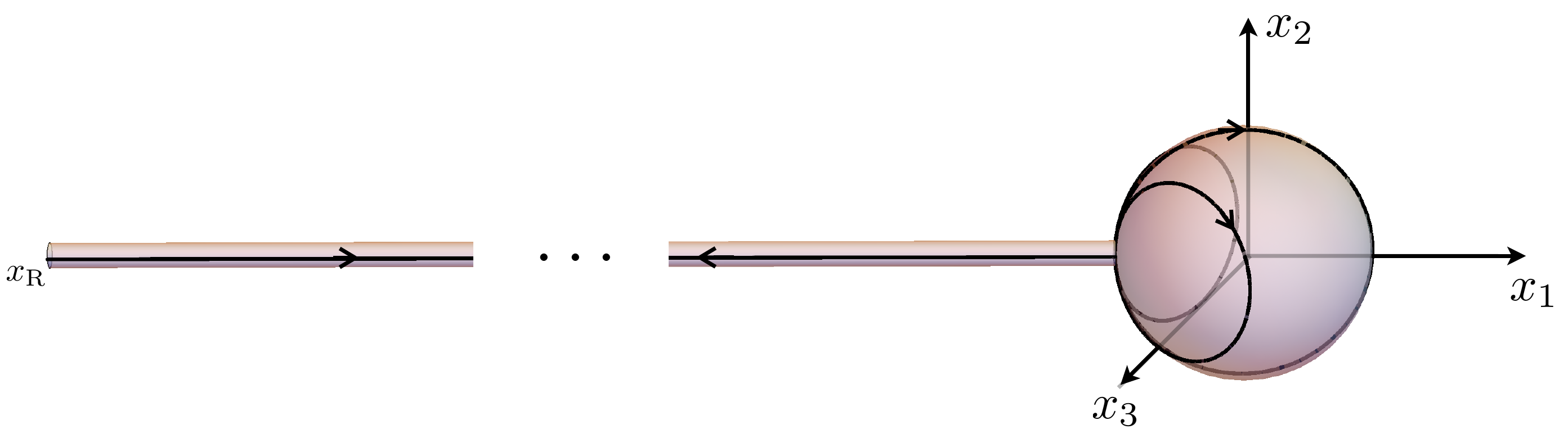}
\caption{Scanning of type (II). }
\label{atwood2}
\end{center}
\end{figure}

 \item {\bf Loops of type (III), scanning the thin cylinder backwards,  as depicted in Figure \ref{atwood3}}. For such loops the parameter $\tau$ varies from  $\frac{\pi}{2}$ to $\infty$, and they are made of two straight lines. The first one starting at $x_R$ and ending on some point on the side of the cylinder with coordinates  $\(x_1\,,\, x_2\,,\,x_3\)=\(x_1\,,\,0\,,\, -\ve\)$. The second part of the loop is the same   straight line (reversed) going back to $x_R$. We shall denote the first straight line (III.1) and the second (III.2). We parameterize the loops with $\sigma$, such that the points on the loops have the following coordinates:
 $$
\begin{array}{lllll}
\textrm{(III.1)} & \qquad x_1 = \frac{\pi}{2}-\tau - \zeta + \sigma & \qquad x_2 = 0 & \qquad x_3 = -\varepsilon & \qquad (-\infty\leq \sigma \leq 0)\\
\textrm{(III.2)} & \qquad x_1 = \frac{\pi}{2}-\tau - \zeta - \sigma & \qquad x_2 = 0 & \qquad x_3 = -\varepsilon & \qquad (0\leq \sigma \leq \infty)
\end{array}
$$
with fixed $\zeta$, and where  $ \frac{\pi}{2} \leq \tau \leq \infty$.

\begin{figure}[htbp]
\begin{center}
\includegraphics[width=0.6\textwidth]{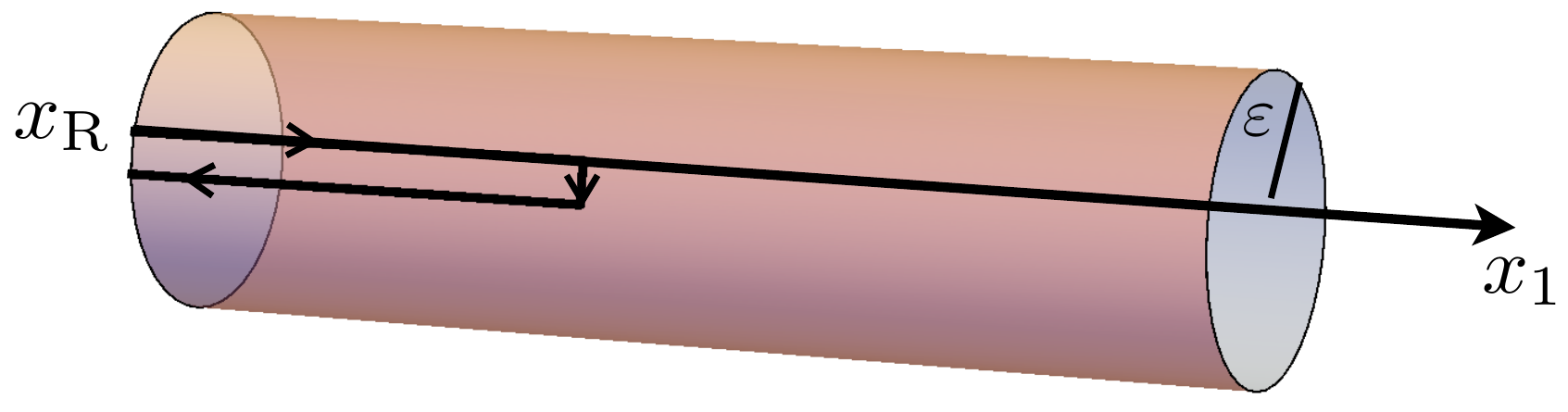}
\caption{Scanning of type (III). }
\label{atwood3}
\end{center}
\end{figure}
 \end{enumerate}

An important simplification is made by observing that the Wilson line is constant along loops scanning the thin cylinder. Indeed, we observe that on the segments (I.1), (I.3), (II.1), (II.3), (III.1) and (III.2), the coordinate $x_1$ is linear in $\sigma$, and $x_2$ and $x_3$ are independent of it. Therefore,  using \rf{bpsmonopole}, we have that
\be
A_i\,\frac{dx^i}{d\sigma}\mid_{{\rm straight}\,{\rm lines}}=\pm  A_1= \mp \frac{1}{e}\,\frac{\ve}{r^2}\,\(1-K\)\,T_2\rightarrow 0
\qquad\qquad {\rm as}\qquad \ve\rightarrow 0
\lab{straightline}
\ee
with the upper signs valid for the segments (I.1), (II.1) and (III.1), and the lower signs valid for (I.3), (II.3) and (III.2). On the segment (I.2), on the other hand, we get that
\br
A_i\,\frac{dx^i}{d\sigma}\mid_{{\rm (I.2)}}&=&\ve\left[\cos \sigma\, A_2+\sin\sigma\, A_3\right]
\\
&=& -\frac{\ve}{e}\,\frac{\(1-K\)}{r^2}\,\left[ -\ve\,T_1 +\(\tau-\zeta+\frac{\pi}{2}\)\(-\cos\sigma\, T_3+\sin\sigma\,T_2\) \right]\rightarrow 0
\qquad\quad {\rm as}\quad \ve\rightarrow 0
\nonumber
\er
The only non-vanishing contribution comes from the segment (II.2) which gives
\br
A_i\frac{dx^i}{d\sigma}\mid_{{\rm (II.2)}}&=&\frac{1}{e}(1-K)\cos\tau \left[\cos\tau\, \sin\tau (1-\cos\sigma)T_1+\sin\tau\,\sin\sigma \,T_2 + \left(\sin^2\tau (1-\cos\sigma)-1\right)T_3\right]
\nonumber\\
&=&-\frac{1}{e}(1-K)\cos\tau\; e^{i\,\tau\,T_2}\,e^{i\,\sigma\,T_3}\,e^{-i\,\tau\,T_2}\,T_3\,
e^{i\,\tau\,T_2}\,e^{-i\,\sigma\,T_3}\,e^{-i\,\tau\,T_2}
\lab{contrib_wilson}
\er
Therefore, integrating \rf{w} one gets that the Wilson lines on the loops of type I and III are trivial, i.e. $W\(I\)= W\(III\)=\one$. On the loops of type II one gets that $W\(II\)=W_3\,W_2\,W_1$, where $W_a$ are the Wilson lines obtained by integrating \rf{w} on the segments (II.a), a$=1,2,3$. Due to \rf{straightline} we have that $W_1=W_3=\one$. Under a gauge transformation $A_i\rightarrow {\bar A}_i= g\, A_i\,g^{-1}+\frac{i}{e}\,\partial_ig\,g^{-1}$, with $g= e^{i\,\tau\,T_2}\,e^{-i\,\sigma\,T_3}\,e^{-i\,\tau\,T_2}$,  one gets that
\be
W_2\rightarrow {\bar W}_2=g_f\, W_2\, g_i ^{-1}= e^{i\,\tau\,T_2}\,e^{-i\,2\,\pi\,T_3}\,e^{-i\,\tau\,T_2}
W_2
\ee
where $g_i$ and $g_f$ are the values of $g$ at the initial and final points of the loop (II.2), and so $g_i=\one$, and $g_f=e^{i\,\tau\,T_2}\,e^{-i\,2\,\pi\,T_3}\,e^{-i\,\tau\,T_2}$. Therefore, one gets that
\be
A_i\frac{dx^i}{d\sigma}\mid_{{\rm (II.2)}}\rightarrow
{\bar A}_i\frac{dx^i}{d\sigma}\mid_{{\rm (II.2)}}=\frac{1}{e}\left[K\, \cos \tau\,T_3-\sin\tau\, T_1\right]
\ee
 and  the equation \rf{w} for ${\bar W}_2$ becomes
\be
\frac{d\,{\bar W}_2}{d\,\sigma} + i\, \left[K\, \cos \tau\,T_3-\sin\tau\, T_1\right]\, {\bar W}_2=0.
\ee
Since the connection term $\left[K\(\zeta\)\, \cos \tau\,T_3-\sin\tau\, T_1\right]$ is independent of $\sigma$ it follows that the path ordering is unimportant and the integration on the loops (II.2) gives
\be
{\bar W}_2= e^{-i\,2\,\pi\, \left[K\, \cos \tau\,T_3-\sin\tau\, T_1\right]}
\ee
Using the fact that $e^{i\,2\,\pi\, T_3}= \pm \one$, depending if the representation used is of integer ($+$) or half-integer ($-$) spin,  we get that $g_f=\pm\,\one$, and so
\be
W=W\(II\)=W_2=\pm \,
e^{-i\,2\,\pi\, \left[K\(\zeta\)\, \cos \tau\,T_3-\sin\tau\, T_1\right]}
\lab{w2quasifinal1}
\ee
where we have equated $W$ to $W\(II\)$, because, as shown above $W\(I\)= W\(III\)=\one$. Therefore, in \rf{w2quasifinal1} we have $\tau$ varying from $-\frac{\pi}{2}$ to $\frac{\pi}{2}$.
The calculations concerning the Wilson line can be simplified defining  $\gamma$ as
\be
\cos\gamma=\frac{K\,\cos\tau}{F}\,;\; \qquad\qquad\qquad\qquad
\sin\gamma=\frac{\sin\tau}{F}
\lab{gammadef}
\ee
with
\be
F\(\zeta\,,\,\tau\)=\sqrt{K^2\(\zeta\)\,\cos^2\tau+\sin^2\tau}.
\lab{fdef}
\ee
Then \rf{w2quasifinal1} can be written as
\be
W = \pm \;
e^{i\,\gamma\, T_2}\, e^{-i\,2\,\pi\, F\,T_3}\,e^{-i\,\gamma\, T_2},
\lab{w2final}
\ee
 from which we get
\br
W^{-1}\,\partial\,W&=& i\, e^{i\,\gamma\, T_2}\, \left\{
- 2\,\pi\,\partial F\, T_3+\partial \gamma\,\left[\(\cos\(2\,\pi\,F\)-1\)\, T_2+\sin\(2\,\pi\,F\)\, T_1\right]
\right\}\,e^{-i\,\gamma\, T_2}
\nonumber\\
&=&i\, \left\{ \left[2\,\pi\,\partial F\, \sin\gamma + \partial \gamma\,\sin\(2\,\pi\,F\)\,\cos \gamma\right]\, T_1
+\partial \gamma\,\left[\cos\(2\,\pi\,F\)-1\right] T_2
\right. \nonumber\\
&+&\left. \left[-2\,\pi\,\partial F\, \cos\gamma + \partial \gamma\,\sin\(2\,\pi\,F\)\,\sin \gamma\right]\,T_3
\right\}.
\er
We then have
\br
W^{-1}\frac{d\,W}{d\,\tau}= i\, \cos \tau\; N_j\(K\,,\,\tau\)\, T_j
\lab{wdtau}
\er
with
\br
N_1\(K\,,\,\tau\)&=&\frac{1}{F^2}\left[2\,\pi\,\(1-K^2\)\,\sin^2\tau+\frac{K^2\,\sin\(2\,\pi\,F\)}{F}\,\right]
\nonumber\\
N_2\(K\,,\,\tau\)&=&-\frac{K}{F^2\,\cos\tau}\left[1-\cos\(2\,\pi\,F\)\right]
\lab{nadef}\\
N_3\(K\,,\,\tau\)&=&\frac{K\,\sin\tau}{F^2\,\cos\tau}\left[-2\,\pi\,\(1-K^2\)\,\cos^2\tau+\frac{\sin\(2\,\pi\,F\)}{F}\right].
\nonumber
\er
In addition
\br
\sbr{W^{-1}\frac{d\,W}{d\,\tau}}{W^{-1}\frac{d\,W}{d\,\zeta}}&=& 2\,\pi\,i\(\frac{d\,F}{d\,\tau}\,\frac{d\,\gamma}{d\,\zeta}-\frac{d\,\gamma}{d\,\tau}\frac{d\,F}{d\,\zeta}\)
\nonumber\\
&\times&
e^{i\,\gamma\, T_2}\, \left[\(1-\cos\(2\,\pi\,F\)\)\, T_1+\sin\(2\,\pi\,F\)\, T_2\right]\,e^{-i\,\gamma\, T_2}
\nonumber\\
&=& -i\, 2\,\pi\,K^{\prime}\,\cos^2\tau\; M_j\(K\,,\,\tau\)\, T_j
\lab{wdtauwdzeta}
\er
with
\br
M_1\(K\,,\,\tau\)&=&\frac{K\,\cos\tau}{F^2}\,\left[1-\cos\(2\,\pi\,F\)\right]
\nonumber\\
M_2\(K\,,\,\tau\)&=&\frac{\sin\(2\,\pi\,F\)}{F}
\lab{mdefwdtauwdzeta}\\
M_3\(K\,,\,\tau\)&=&\frac{\sin\tau}{F^2}\,\left[1-\cos\(2\,\pi\,F\)\right]
\nonumber
\er
where  $K'$ stands for $\frac{dK}{d\zeta}$, and where we have used the formulas
\be
\frac{d\,F}{d\,\tau} = \frac{\sin\tau\,\cos\tau}{F}\(1-K^2\) \, ,
\;\;\;
\frac{d\,F}{d\,\zeta} = \frac{K\,\cos^2\tau}{F}\, K^{\prime}\, ,
\;\;\;
\frac{d\,\gamma}{d\,\tau}=\frac{K}{F^2}\, ,
\;\;\;
\frac{d\,\gamma}{d\,\zeta}= - \frac{\sin\tau\,\cos\tau}{F^2}\,K^{\prime}.
\lab{derivatives}
\ee
With  these expressions we are ready to perform the calculations of section \ref{sec:expansion} for the $SU(2)$  monopoles.

\section{Check of first order integral equations  for $SU(2)$ monopoles}
\label{sec:firstorder}
\setcounter{equation}{0}

The integral equation for a purely spatial volume $\Omega$, in first order in $\alpha$, for the $SU(2)$ monopoles ('t Hooft-Polyakov or BPS) is given by expression  \rf{firstorderbianchiint}. However, since the Wilson line is unit for the loops of type I and III (see section \ref{sec:wilson}) we get that   \rf{firstorderbianchiint} is only non-trivial for loops of type II, where $\tau$ varies from $-\frac{\pi}{2}$ to $\frac{\pi}{2}$, and so \rf{firstorderbianchiint} becomes
\begin{equation}
\lab{eq1a}
V_{(1)}=\int_{-\frac{\pi}{2}}^{\frac{\pi}{2}}d\tau \;W^{-1}\frac{dW}{d\tau}\Bigg\vert_{\zeta = \zeta_0} = \int_0^{\zeta_0}d\zeta \int_{-\frac{\pi}{2}}^{\frac{\pi}{2}}d\tau\;\left[W^{-1}\frac{dW}{d\tau},W^{-1}\frac{dW}{d\zeta}\right]=U_{(1)}
\end{equation}
where the l.h.s is an integration on a closed surface of radius $\zeta_0$ and the r.h.s is an integration in the volume contained inside that surface.
Our goal is to evaluate both sides of this equation using the results obtained in the expressions  \rf{wdtau} and \rf{wdtauwdzeta}.
In order to perform the integration of the l.h.s term, a better choice of variables is the following:
\begin{equation}
y = \sin\tau\;; \qquad  -1 \leq y \leq 1 \;; \qquad \qquad\qquad
z = K(\zeta)\cos\tau\;;  \qquad\qquad  0 \leq z \leq 1
\lab{param1}
\end{equation}
with $0\leq \zeta \leq \infty$, $0\leq K(\zeta)\leq 1$, and $-\frac{\pi}{2}\leq \tau \leq \frac{\pi}{2}$. Note that we are using the fact that $K(\zeta)$ is monotonically decreasing function of $\zeta$ for both, the 't Hooft-Polyakov and BPS monopole solutions. The explicit form of the function $K(\zeta)$ is not important here.  In these variables we get
\begin{equation}
F^2 = y^2+z^2 = K^2+(1-K^2)y^2
\end{equation}
 and so using \rf{wdtau} and \rf{nadef} we get that the l.h.s. of \rf{eq1a} becomes
 \begin{equation}
 V_{(1)}=i\,\int_{-1}^1dy\, N_j\(K_0\,,\,y\)\,T_j
\lab{dtauw}
\end{equation}
with $K_0 \equiv K(\zeta_0)$, and
\begin{eqnarray}
\lab{niydef}
N_1(K\,,\,y)&=&\frac{2\pi}{F^2}\left\{ y^2 (1-K^2)+\frac{K^2 \sin(2\pi F)}{2\pi F}\right\} \nonumber\\
N_2(K\,,\,y)&=&-\frac{K}{\sqrt{1-y^2}F^2}\left\{1-\cos{(2\pi F}\right\}\\
N_3(K\,,\,y)&=&2\pi\frac{K\, y}{\sqrt{1-y^2}F^2}\left\{ F^2 - 1 + \frac{\sin{(2\pi F)}}{2\pi F}\right\}\nonumber.
\end{eqnarray}

 Note that $N_3$ is an odd function of $y$ and so integrating we get
\begin{equation}
\label{eq1alhs}
V_{(1)}(K_0)= i\,J_1(K_0)T_1 + i\,J_2(K_0)T_2,
\end{equation}
with
\begin{eqnarray*}
J_1(K_0) &=& 2\pi \int_{-1}^{1}dy\; \frac{1}{K_0^2 + (1-K_0^2)y^2}\left\{ (1-K_0^2)y^2 + \frac{K_0^2 \sin\left(2\pi \sqrt{K_0^2 + (1-K_0^2)y^2}\right)}{2\pi \sqrt{K_0^2 + (1-K_0^2)y^2}}\right\}\\
J_2(K_0) &=& -\int_{-1}^{1}dy\;\frac{K_0}{\sqrt{1-y^2}\left(K_0^2 + (1-K_0^2)y^2\right)}\left\{1-\cos{\left( 2\pi\sqrt{K_0^2 + (1-K_0^2)y^2} \right)}\right\}.
\end{eqnarray*}

\begin{figure}[h!]
  \begin{center}
  \includegraphics[width=9cm, height=9cm, keepaspectratio]{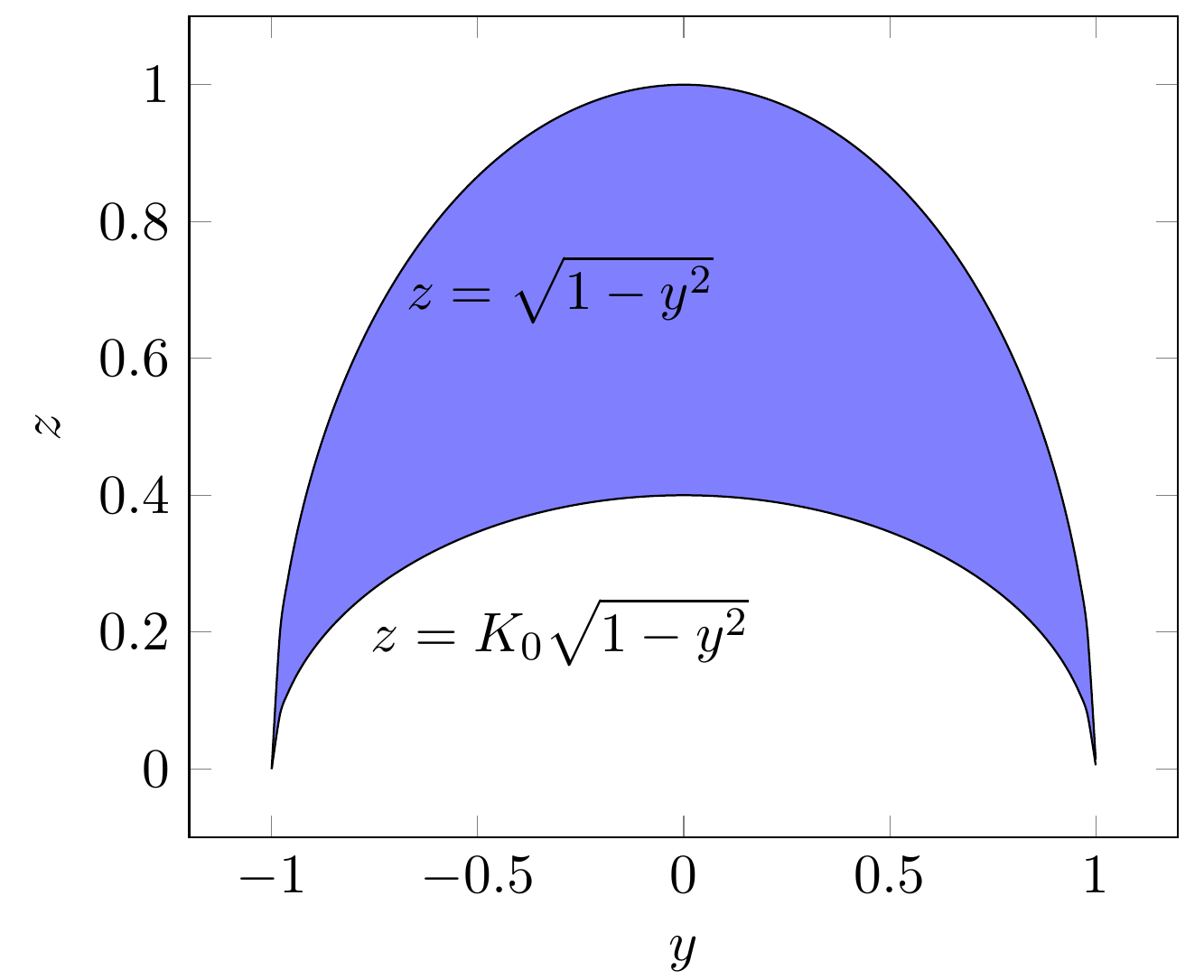}
  \caption{The integration domain in the new ``polar'' coordinates. Each value of $K_0$ fixes a new domain by shortening the area of the disk from below.}
\label{fig:semidisc}
\end{center}
\end{figure}

Note that as $\zeta$ varies from $0$ to $\zeta_0$, one has that $K$ varies from $1$ to $K_0\equiv K\(\zeta_0\)< 1$. Therefore, the integration domain on the r.h.s. of \rf{eq1a} is a truncated semi-disc shown in Figure \ref{fig:semidisc}. The absolute value of the Jacobian of the variable transformation $\(\zeta\,,\,\tau\)\rightarrow \(y\,,\,z\)$, given in \rf{param1}, is $\mid K^{\prime}\mid \cos^2\tau= - K^{\prime} \cos^2\tau$,  since $K^{\prime}$ is strictly negative. In addition, it is more appropriate to perform a further change of variables to evaluate the integration on the r.h.s. of \rf{eq1a}. We define the polar type coordinates $\(s\,,\, \theta\)$ as
\be
y=s\cos\theta\;;\qquad\qquad\qquad z=s\sin\theta\;; \qquad\qquad S\(K_0\,,\,\theta\)\leq s\leq 1\;;\qquad\qquad 0 \leq\theta\leq \pi
\lab{sthetadef}
\ee
with
\be
S(K\,,\,\theta) \equiv  \frac{K}{\sqrt{1-\cos^2\theta(1-K^2)}}
\lab{skthetadef}
\ee
 Therefore one has that
\be
\label{tsemidisc}
\int_0^{\zeta_0}d \zeta \int_{-\frac{\pi}{2}}^{\frac{\pi}{2}} d\tau K^\prime \cos^2\tau= -\int_{\mbox{truncated semi-disc}}dz dy = - \int_0^{\pi} d\theta\int_{S(K_0,\theta)}^1 ds\, s
\ee

We then have that  the $M_i$'s, defined in \rf{mdefwdtauwdzeta},  become
\br
M_1&=&\frac{z}{F^2}\,\left[1-\cos\(2\,\pi\,F\)\right]
=\frac{\sin\theta}{s}\,\left[1-\cos\(2\,\pi\,s\)\right]
\nonumber\\
M_2&=&\frac{\sin\(2\,\pi\,F\)}{F}=\frac{\sin\(2\,\pi\,s\)}{s}
\lab{mdefwdtauwdzeta2}\\
M_3&=&\frac{y}{F^2}\,\left[1-\cos\(2\,\pi\,F\)\right]
=\frac{\cos\theta}{s}\,\left[1-\cos\(2\,\pi\,s\)\right]
\nonumber
\er
Using \rf{wdtauwdzeta} we get that in these coordinates the r.h.s. of \rf{eq1a}, denoted by $U_{(1)}$, reads
\be
U_{(1)}=i2\pi \int_0^{\pi}d\theta \int_{S\(K_0,\theta\)}^1 ds\left\{ \sin\theta (1-\cos(2\pi s))T_1 + \sin(2\pi s)T_2 + \cos\theta (1-\cos(2\pi s))T_3\right\}
\lab{u(1)integral}
\ee
from where we can easily perform the integration in $s$, obtaining
\begin{equation}
\lab{eq1arhs}
U_{(1)}(K_0)= iI_1(K_0)T_1 + iI_2(K_0)T_2
\end{equation}
where
\begin{eqnarray}
\label{i1i2}
I_1(K_0) &=& \int_0^\pi d\theta \sin\theta \left\{ 2\pi - \frac{2\pi K_0}{\sqrt{1-\cos^2{\theta}(1-K_0^2)}}+\sin{\left( \frac{2\pi K_0}{\sqrt{1-\cos^2{\theta}(1-K_0^2)}} \right)}\right\} \nonumber\\
I_2(K_0) &=& - \int_0^\pi d\theta \left\{ 1 - \cos{\left(\frac{2\pi K_0}{\sqrt{1-\cos^2{\theta}(1-K_0^2)}}\right)}\right\}
\end{eqnarray}
The integral along the $T_3$-direction  in \rf{u(1)integral} vanishes since the integrand is odd, under reflection around $\theta=\frac{\pi}{2}$,  in the interval $0 \leq\theta\leq \pi$ (note that $S_0\(\theta\)$ is even in that interval).

Therefore, in order to check the validity of the integral equation at first order in $\alpha$, given in \rf{eq1a}, we have to verify the equalities of the coefficients of $T_i$ in \eqref{eq1alhs} and in  \rf{eq1arhs}. We have performed the numerical integration of the quantities   $I_i(K_0)$ and $J_i(K_0)$  for several values of $K_0$, covering the range $1\geq K_0\geq 0$, corresponding to $0\leq \zeta_0\leq \infty$.  Note that the actual value of $K_0$ for a given value of $\zeta_0$ is different for the 't Hooft-Polyakov and BPS monopoles. However, the fact that $K\(\zeta\)$ is a monotonically decreasing function of $\zeta$, for both solutions, allowed us to trade the coordinate $\zeta$ by $K$, and perform one check that is valid for the two monopole solutions. In section  \ref{sec:tablefirstorder} we give the results of the numerical integrations of the quantities $I_i(K_0)$ and $J_i(K_0)$,  $i=1,2$.
 As one observes in those tables  the values of $I_i(K_0)$ and $J_i(K_0)$ are remarkably identical, differing in the worst case around the eighth decimal place, due to the numerical approximation. This indicates that the 't Hooft-Polyakov and BPS $SU(2)$ monopoles are indeed  solutions of the first order integral equation \rf{orderalpha}, or equivalently \rf{eq1a}, appearing in the expansion in $\alpha$ of the integral non-abelian Gauss law in \rf{lhsbianchi} and \rf{rhsbianchi}.

\section{Check of second order integral equations  for $SU(2)$ monopoles}
\label{sec:secondorder}
\setcounter{equation}{0}

The integral equation for a purely spatial volume $\Omega$, in second order in $\alpha$, for the $SU(2)$ monopoles ('t Hooft-Polyakov or BPS) is given by expression  \rf{secondorderbianchiint}. However, since the Wilson line is unit for the loops of type I and III (see section \ref{sec:wilson}) we get that   \rf{secondorderbianchiint} is only non-trivial for loops of type II, where $\tau$ varies from $-\frac{\pi}{2}$ to $\frac{\pi}{2}$, and so \rf{secondorderbianchiint}  becomes
\begin{eqnarray}
\lab{eq2a}
V_{(2)}&=&\int_{-\frac{\pi}{2}}^{\frac{\pi}{2}}d\tau\int_{-\frac{\pi}{2}}^{\tau}d\tau^{\prime}\;W^{-1}\frac{dW}{d\tau^{\prime}}W^{-1}\frac{dW}{d\tau}\Bigg\vert_{\zeta = \zeta_0} = -\int_0^{\zeta_0}d\zeta \int_{-\frac{\pi}{2}}^{\frac{\pi}{2}}d\tau\; \left[W^{-1}\frac{dW}{d\tau},W^{-1}\frac{dW}{d\zeta}\right] \nonumber\\
&+&\int_0^{\zeta_0}d\zeta\int_{-\frac{\pi}{2}}^{\frac{\pi}{2}}d\tau\int_{-\frac{\pi}{2}}^\tau d\tau'\left[ W^{-1}\frac{dW}{d\tau'},\left[ W^{-1}\frac{dW}{d\tau},W^{-1}\frac{dW}{d\zeta}\right]\right]\\
&+&\int_0^{\zeta_0}d\zeta \int_0^\zeta d\zeta' \int_{- \frac{\pi}{2}}^{\frac{\pi}{2}}d\tau\int_{- \frac{\pi}{2}}^{\frac{\pi}{2}}d\tau'\;\left[ W^{-1}\frac{dW}{d\tau},W^{-1}\frac{dW}{d\zeta}\right]\left[ W^{-1}\frac{dW}{d\tau'},W^{-1}\frac{dW}{d\zeta'}\right]
\nonumber\\
&\equiv& - U_{(1)} + G_2+G_3= U_{(2)}
\nonumber
\end{eqnarray}
where we have denoted $G_2$ and $G_3$ the  terms appearing on the second and third lines respectively of \rf{eq2a}. In addition, we have used the fact that the first term on r.h.s. of the first line of \rf{eq2a} is the same (up to a minus sign) as  $U_{(1)}$ given on the r.h.s. of \rf{eq1a}.

We start by evaluating the l.h.s. of \rf{eq2a}, using \rf{wdtau}, and \rf{niydef}   to get
\br
&&\int_{-\frac{\pi}{2}}^{\frac{\pi}{2}}d\tau\int_{-\frac{\pi}{2}}^{\tau}d\tau^{\prime} \;W^{-1}\frac{dW}{d\tau^{\prime}}W^{-1}\frac{dW}{d\tau}\Bigg\vert_{\zeta = \zeta_0}=-\int_{-1}^1dy\,\int_{-1}^ydy^{\prime}\, \sum_{i,j=1}^3 \, N_i\(K_0\,,\, y^{\prime}\)\,N_{j}\(K_0\,,\, y\)\, T_i\,T_j
\nonumber\\
&=&-\frac{1}{2}\sum_{i=1}^3\left[\int_{-1}^1dy\,N_i\(K_0\,,\, y\)\right]^2\,T_i^2 -
\int_{-1}^1dy\,\int_{-1}^ydy^{\prime}\, \sum_{i\neq j=1}^3 \, N_i\(K_0\,,\, y^{\prime}\)\,N_{j}\(K_0\,,\, y\)\, T_i\,T_j
\lab{firsttermpre}
\er
where in the first term on the r.h.s. of \rf{firsttermpre} we have used the symmetry of the integrand in $y$ and $y^{\prime}$ to transform the integral on the triangle $-1\leq y\leq 1$ and $y^{\prime}\leq y$, to the integral on the square $-1\leq y\,,\, y^{\prime}\leq 1$. We now use the fact that $N_i\(K_0\,,\, -y\)=\ve_i\,N_i\(K_0\,,\, y\)$, with $\ve_i=1$ for $i=1,2$ and $\ve_3=-1$ (see \rf{niydef}). Then we can write
\br
\int_{-1}^1dy\,\int_{-1}^ydy^{\prime}\,  N_i\(K_0\,,\, y^{\prime}\)\,N_{j}\(K_0\,,\, y\)&=&
\frac{1}{2}\,\int_{-1}^1dy\,\int_{-1}^ydy^{\prime}\,  N_i\(K_0\,,\, y^{\prime}\)\,N_{j}\(K_0\,,\, y\) \nonumber\\
&+&
\frac{\ve_i\,\ve_j}{2}\,\int_{-1}^1dy\,\int_{y}^1dy^{\prime}\,  N_i\(K_0\,,\, y^{\prime}\)\,N_{j}\(K_0\,,\, y\)
\lab{nicesplitting}
\er
Therefore for the case where $\ve_i\,\ve_j=1$, one can write further that
\br
\int_{-1}^1dy\,\int_{-1}^ydy^{\prime}\,  N_i\(K_0\,,\, y^{\prime}\)\,N_{j}\(K_0\,,\, y\)&=&
\frac{1}{2}\,\int_{-1}^1dy\,\int_{-1}^1dy^{\prime}\,  N_i\(K_0\,,\, y^{\prime}\)\,N_{j}\(K_0\,,\, y\)\;;\qquad
\ve_i\,\ve_j=1
\nonumber
\er
For the case $\ve_i\,\ve_j=-1$, we do not use \rf{nicesplitting}, but instead write
\be
T_i\,T_j=\frac{1}{2}\,\{T_i,T_j\}+ \frac{1}{2}\,\sbr{T_i}{T_j}=\frac{1}{2}\,\{T_i,T_j\}+i\,\ve_{ijk}\,T_k
\lab{producttitj}
\ee
Note that we are dealing here with products, and not only commutators, of the $SU(2)$ Lie algebra generators. We have therefore to work with a basis in the enveloping
 algebra of $SU(2)$, which in the case of quadratic terms we shall take to be the $9$ quantities $T_i$,  and the anti-commutators  $\{T_i,T_j\}$, $i,j=1,2,3$. If one works with the spinor representation given by the Pauli matrices $\sigma_i$, $i=1,2,3$, then one has $\sigma_i\,\sigma_j=i\,\ve_{ijk}\,\sigma_k+ \delta_{ij}\,\one$, and non-diagonal terms vanish, i.e. $\{\sigma_i,\sigma_j\}=0$, for $i\neq j$. However, if one works with the triplet or higher representations one has $\{T_i,T_j\}\neq 0$ even for $i\neq j$. So, we have to consider the coefficients of all the $9$ elements of the basis of the enveloping algebra to be independent. Therefore, using \rf{producttitj} one gets that
\br
V_{(2)}&=&\int_{-\frac{\pi}{2}}^{\frac{\pi}{2}}d\tau\int_{-\frac{\pi}{2}}^{\tau}d\tau^{\prime} \;W^{-1}\frac{dW}{d\tau^{\prime}}W^{-1}\frac{dW}{d\tau}\Bigg\vert_{\zeta = \zeta_0}=
-\left[ \mathcal{N}_1(K_0)T_1^2 + \mathcal{N}_2(K_0)T_2^2 + \mathcal{N}_{12}(K_0)\left\{T_1,T_2\right\}
\right. \nonumber\\
&+& \left.\mathcal{N}_{13}^{+}(K_0)\left\{T_1,T_3\right\}+\mathcal{N}_{23}^{+}(K_0)\left\{T_2,T_3\right\}
- i\,\mathcal{N}_{13}^{-}(K_0)\,T_2+i\,\mathcal{N}_{23}^{-}(K_0)\,T_1\right]
\lab{v2final}
\er
where
\br
\mathcal{N}_i(K_0) &=& \frac{1}{2}\left( \int_{-1}^{1}dy\;N_i(K_0\,,\, y)\right)^2 \qquad i = 1,2
\nonumber\\
\mathcal{N}_{12}(K_0) &=& \frac{1}{2}\left(\int_{-1}^{1}dy\;N_1(K_0\,,\, y)\right)\left(\int_{-1}^1 dy'N_2(K_0\,,\, y')\right)
\\
\mathcal{N}_{13}^{\pm}(K_0)&=&\frac{1}{2}\left(\int_{-1}^{1}dy\int_{-1}^y N_1(K_0\,,\, y')N_3(K_0\,,\, y)\pm  \int_{-1}^{1}dy\int_{-1}^{y}dy'N_3(K_0\,,\, y')N_1(K_0\,,\, y)\right)
\nonumber\\
\mathcal{N}_{23}^{\pm}(K_0) &=& \frac{1}{2}\left(\int_{-1}^{1}dy\int_{-1}^y dy'N_2(K_0\,,\, y')N_3(K_0\,,\, y)\pm \int_{-1}^{1}dy \int_{-1}^y dy'N_3(K_0\,,\, y')N_2(K_0\,,\, y)\right)
\nonumber
\er
with the $N_i$'s defined in \rf{niydef}, and where we have dropped the term proportional to $T_3^2$ because $N_3$ is an odd function of $y$, and so its integral on the interval $-1\leq y\leq 1$, vanishes.

Using \rf{wdtau}, \rf{wdtauwdzeta} and \rf{param1}, the term on the second line of \rf{eq2a}, denoted $G_2$, becomes
\br
G_2&=& -i\,2\,\pi\,  \ve_{ijk}\,T_k\,\int_1^{K_0}dK\,\int_{-1}^1dy\,\sqrt{1-y^2}\,\, M_j\(K\,,\,y\)\int_{-1}^y\,dy^{\prime}\, N_i\(K\,,\,y^{\prime}\)
\nonumber\\
&\equiv& -i\,4\,\pi^2\, R_k\(K_0\)\, T_k
\lab{g2final}
\er
Using \rf{wdtauwdzeta} and \rf{mdefwdtauwdzeta2} the term on the third line of \rf{eq2a}, denoted $G_3$, becomes
\br
G_3=-4\,\pi^2\,\int_0^{\pi}d\theta\,\int_{S(K_0,\theta)}^1ds\,s\,\int_0^{\pi}d\theta^{\prime}\,\int_{S(K,\theta^{\prime})}^1ds^{\prime}\, s^{\prime}\,\sum_{i,j=1}^3 M_i\(s\,,\,\theta\)\,M_j\(s^{\prime}\,,\,\theta^{\prime}\)\, T_i\,T_j
\er
with $K\geq K_0$, and so $\zeta\leq \zeta_0$. Note that in the $\(\theta^{\prime}\,,\,s^{\prime}\)$-integration, $K$ has to be taken as a function of $\theta$ and $s$. From \rf{param1} and \rf{sthetadef} one gets that $K= \frac{s\,\sin\theta}{\sqrt{1-s^2\,\cos^2\theta}}$. Note that the $\(\theta^{\prime}\,,\,s^{\prime}\)$-integration is the same as the one performed in \rf{u(1)integral}, with $K_0$ replaced by $K$. Therefore, similar to what happened,  there will be no terms in the direction of $T_j$ for $j=3$, since $M_3\(s^{\prime}\,,\,\theta^{\prime}\)$ is odd under reflection of $\theta^{\prime}$ around $\theta^{\prime}=\frac{\pi}{2}$ (see \rf{mdefwdtauwdzeta2}). Since $K$ and $S\(K_0,\theta\)$ are even under the reflection of $\theta$ around $\theta=\frac{\pi}{2}$, there will be no terms in the direction of $T_i$ for $i=3$, since $M_3\(s\,,\,\theta\)$ is odd under that reflection. Using \rf{producttitj}  one gets that
\begin{eqnarray*}
G_3 &=& -4\pi \left[ S_1(K_0)T_1^2 + S_2(K_0)T_2^2 + S_{12}\left\{T_1,T_2\right\}+iS_3(K_0)T_3\right]
\lab{g3final}
\end{eqnarray*}
with
\br
S_a(K_0)&=&\pi\,\int_0^{\pi}d\theta\,\int_{S(K_0,\theta)}^1ds\,s\,\int_0^{\pi}d\theta^{\prime}\,\int_{S(K,\theta^{\prime})}^1ds^{\prime}\, s^{\prime}\, M_a\(s\,,\,\theta\)\,M_a\(s^{\prime}\,,\,\theta^{\prime}\)\;;\qquad a=1,2
\nonumber\\
S_{12}(K_0)&=&\frac{\pi}{2}\,\int_0^{\pi}d\theta\,\int_{S(K_0,\theta)}^1ds\,s\,\int_0^{\pi}d\theta^{\prime}\,\int_{S(K,\theta^{\prime})}^1ds^{\prime}\, s^{\prime}\,
\nonumber\\
&\times& \left[M_1\(s\,,\,\theta\)\,M_2\(s^{\prime}\,,\,\theta^{\prime}\)+ M_2\(s\,,\,\theta\)\,M_1\(s^{\prime}\,,\,\theta^{\prime}\)\right]
\\
S_{3}(K_0)&=&\frac{\pi}{2}\,\int_0^{\pi}d\theta\,\int_{S(K_0,\theta)}^1ds\,s\,\int_0^{\pi}d\theta^{\prime}\,\int_{S(K,\theta^{\prime})}^1ds^{\prime}\, s^{\prime}\,
\nonumber\\
&\times& \left[M_1\(s\,,\,\theta\)\,M_2\(s^{\prime}\,,\,\theta^{\prime}\)- M_2\(s\,,\,\theta\)\,M_1\(s^{\prime}\,,\,\theta^{\prime}\)\right]
\nonumber
\er
The $s^{\prime}$-integration can be performed analytically and so, using \rf{mdefwdtauwdzeta2} and the fact that $K= \frac{s\,\sin\theta}{\sqrt{1-s^2\,\cos^2\theta}}$, we get
\br
\int_{S(K,\theta^{\prime})}^1ds^{\prime}\, s^{\prime}\, M_1\(s^{\prime}\,,\,\theta^{\prime}\)&=&
\sin\theta^{\prime}\left[1-\frac{s\,\sin\theta}{\sqrt{s^2\,\sin^2\theta+\(1-s^2\)\sin^2\theta^{\prime}}}
\right.\nonumber\\
&-&\left. \frac{1}{2\pi}\sin\(\frac{2\,\pi\,s\,\sin\theta}{\sqrt{s^2\,\sin^2\theta+\(1-s^2\)\sin^2\theta^{\prime}}}\)
\right]
\er
and
\br
\int_{S(K,\theta^{\prime})}^1ds^{\prime}\, s^{\prime}\, M_2\(s^{\prime}\,,\,\theta^{\prime}\)&=&
\frac{1}{2\pi}\left[-1+\cos\(\frac{2\,\pi\,s\,\sin\theta}{\sqrt{s^2\,\sin^2\theta+\(1-s^2\)\sin^2\theta^{\prime}}}\)
\right]
\er
Note that the above integrals are symmetric under the reflection of $\theta$ and $\theta^{\prime}$ around $\frac{\pi}{2}$. The quantities $M_1\(s\,,\, \theta\)$, $M_2\(s\,,\, \theta\)$, $S\(K\,,\,\theta\)$ and $K\(s\,,\,\theta\)$ are also symmetric under the reflection of $\theta$ around $\frac{\pi}{2}$. Therefore, the integration in $\theta$ and $\theta^{\prime}$ can be performed in the interval from zero to $\frac{\pi}{2}$, by multiplying the result by two. So, we then get that
\br
S_1(K_0)&=&2\,\int_0^{\frac{\pi}{2}}d\theta \int_{S(K_0,\theta)}^1 ds  \int_0^{\frac{\pi}{2}} d\theta^{\prime} \,\left(1-\cos{(2\pi s)}\right)\,\sin{\theta}\,\sin\theta^{\prime}
\nonumber\\
&\times&\left[2 \pi  - \frac{2 \pi  s \sin \theta}{\sqrt{s^2 \sin^2\theta +(1-s^2)\sin^2\theta^{\prime}}}+   \sin \left( \frac{2 \pi  s \sin \theta}{\sqrt{s^2 \sin^2\theta +(1-s^2)\sin^2\theta^{\prime}}}   \right) \right]
\\
P_{12}(K_0)&=&2\,\int_0^{\frac{\pi}{2}}d\theta \int_{S(K_0,\theta)}^{1}ds \int_{0}^{\frac{\pi}{2}}d\theta^{\prime} \; \left(1-\cos{(2\pi s)}\right)\,\sin\theta\,\sin \theta^{\prime}
\nonumber\\
&\times& \left[  -1 + \cos\left(  \frac{2 \pi  s \sin \theta}{\sqrt{s^2 \sin^2\theta +(1-s^2)\sin^2\theta^{\prime}}} \right) \right]\\
P_{21}(K_0)&=& 2\,\int_0^{\frac{\pi}{2}}d\theta \int_{S_0(\theta)}^1 ds  \int_0^{\frac{\pi}{2}} d\theta^{\prime} \sin\(2 \pi s\)
\nonumber\\
&\times&\left[2 \pi  - \frac{2 \pi  s \sin \theta}{\sqrt{s^2 \sin^2\theta +(1-s^2)\sin^2\theta^{\prime}}} +  \sin \left( \frac{2 \pi  s \sin \theta}{\sqrt{s^2 \sin^2\theta +(1-s^2)\sin^2\theta^{\prime}}}   \right) \right]
\\
S_2(K_0)&=& 2\,\int_0^{\frac{\pi}{2}}d\theta \int_{S(K_0,\theta)}^{1}ds \int_{0}^{\frac{\pi}{2}}d\theta^{\prime} \;\sin{(2\pi s)} \left[  -1 + \cos\left(  \frac{2 \pi  s \sin \theta}{\sqrt{s^2 \sin^2\theta +(1-s^2)\sin^2\theta^{\prime}}} \right) \right]
\nonumber\\
\er
where we have introduced
$$
S_{12}=\frac{1}{2}\left(P_{12}(K_0)+P_{21}(K_0)\right) \qquad S_3 = \frac{1}{2}\left(P_{12}(K_0)-P_{21}(K_0)\right),
$$

Summarizing, we have obtained both sides of  the integral equation in second order in $\alpha$ for a given $K_0$, given in  \rf{eq2a}. From \rf{v2final} we have that
\begin{equation}
  \label{lhs2order}
V_{(2)}=-i\mathcal{N}_{23}^{-}T_1 + i\mathcal{N}_{13}^{-}T_2 -\mathcal{N}_1T_1^2 -\mathcal{N}_2T_2^2 - \mathcal{N}_{12}\left\{T_1,T_2\right\}- \mathcal{N}_{13}^{+}\left\{T_1,T_3\right\}- \mathcal{N}_{23}^{+}\left\{T_2,T_3\right\}
\end{equation}
and, from \rf{eq2a}, \rf{eq1arhs}, \rf{g2final} and \rf{g3final} we have that
\begin{equation}
  \label{rhs2order}
U_{(2)}=-i(I_1+4\pi^2R_1)T_1 + -i(I_2+4\pi^2R_2)T_2-i(4\pi^2 R_3 + 4\pi S_3)T_3 -4\pi S_1 T_1^2 -4\pi S_2 T_2^2 - 4\pi S_{12}\left\{T_1,T_2\right\}.
\end{equation}
We have to check the equality between the coefficients of each element of the basis of the $SU(2)$ enveloping algebra on the expansion of  $V_{(2)}$ and $U_{(2)}$. Those coefficients involve integrals which are calculated numerically for a set of values of $K_0$. The results are presented in the tables of section \ref{sec:tablesecondorder} in the Appendix. The consistency is remarkable and with that check we can state clearly that the the 't Hooft-Polyakov monopole and its BPS version satisfy the integral Yang-Mills equations  up to second order in $\alpha$.

\section{Conclusions}
\label{sec:conclusions}
\setcounter{equation}{0}

The integral Yang-Mills equations appeared from an attempt to understand integrability in higher dimensional space-times \cite{ym1,ym2}.
Through a loop space formulation  \cite{afs1,afs2} one can construct a suitable generalization of the non-abelian Stokes theorem for two-form fields that  can be used naturally to define  conservation laws, thus mimicking the so-called zero curvature representation of integrable field theories in $(1+1)$-dimensions. That has led us to consider the applications of such  non-abelian Stokes theorem to construct the  integral equations for non-abelian gauge theories, generalizing the well known  abelian version of such integral equations used to describe the laws of  electrodynamics.
That was indeed possible, as we have shown in \cite{ym1,ym2}, and the usual differential Yang-Mills equations are obtained from these integral equations when the appropriate limit is taken.

The present paper shows that there is more to be explored. The integral Yang-Mills equations allow the introduction of two $c$-numbers as parameters which arise naturally in the construction of the equations, and as being non-linear, produce a quite non-trivial dependence on those parameters of the surface and volume ordered integrals appearing on both sides of the equation.

We have tested the assumption that the integral Yang-Mills equations are in fact a collection of an infinite number of equations, each one corresponding to the coefficients of the above mentioned expansion in powers of those parameters. This was done by considering the fact that, by construction, a solution of the differential Yang-Mills equation is also a solution of the integral Yang-Mills equation. Thus, using the 't Hooft-Polyakov and BPS monopoles as such configurations, we tested the validity of the equations arising at first and second order in the parameter expansion of the integral Yang-Mills equation. Despite the quite different structures of the terms resulting from the surface and volume ordered integrals, we have checked their equalities with a high numerical precision of at least one part in $10^7$.  In addition, much of the check has been done analytically, and we have obtained an exact expression for the Wilson line operator, on each loop scanning the surfaces and  volumes,  for the $SU(2)$ 't Hooft-Polyakov monopole solution and   its  BPS version (see \rf{w2final}). That result can certainly be useful in many other applications.

The fact that those configurations are solutions of both of the highly non-trivial equations at each order of the expansion, indicates that the parameters could indeed be arbitrary.
The arbitrariness of the parameters leads to a variety of important consequences which can now be considered, such as their role in the conserved charges that arise dynamically from the integral equations and the significance of having an infinite number of integral equations.

\vspace{1cm}

\noindent{\bf Aknowkedgments:} The authors are grateful to partial financial   support by FAPES under contract number 0447/2015. LAF is partially supported by CNPq-Brazil.

\newpage

\appendix

\section{Numerical Results}
\label{app:numerics}
\setcounter{equation}{0}

In this section we show the results of  the numerical integrations related to the terms on the l.h.s and r.h.s of the expansion of the integral equation performed at first and second order in $\alpha$.  The coefficients of the generators of the algebra (eventually, up to a common factor of $i\equiv \sqrt{-1}$) are compared for different values of $K_0$. For each integral estimative there is an associated upper bound on the error, which we represent by using the following notation: $1.372\pm 0.008 \equiv 1.37(2\pm8)$.

\subsection{Equation $V_{(1)} = U_{(1)}$}
\label{sec:tablefirstorder}

\begin{table}[h]
\centering
\begin{tabular}{ |p{0.9cm}||p{4.5cm}|p{4.5cm}|  }
 \hline
 \multicolumn{3}{|c|}{\textbf{Coefficients of} $T_1$} \\
 \hline
 $K_0$ & $ I_1(K_0)$  & $J_1(K_0)$ \\
 \hline
 $ 0.01$ & $12.5614010(8 \pm 2)$ &$ 12.561401086$ \\ \hline
 0.1 &$ 12.077187419(9 \pm 6)$& 12.0771874199\\ \hline
 0.2 & $10.70071291(6 \pm 2) $& 10.7007129168\\ \hline
 0.3 & $8.6878758(4\pm 8)  $  & 8.68787584542\\ \hline
 0.4 &$ 6.38863592(8\pm 4) $  & 6.38863592858\\ \hline
 0.5 & 4.169079306          & 4.169079306\\ \hline
 0.6 &$ 2.3285155680(5\pm 6) $& 2.32851556805\\ \hline
 0.7 & $1.0380042(9\pm 1) $   & 1.0380042978\\ \hline
 0.8 & $0.3151104413(2\pm 9)$ & 0.315110441326\\ \hline
 0.9 & 0.039159443823       & 0.039159443823\\ \hline
 0.99 &$ 0.000037982(0\pm 3)$ &$ 0.00003798206260(6\pm 1)$\\ \hline
\end{tabular}

\begin{tabular}{ |p{0.9cm}||p{4.5cm}|p{4.5cm}|  }
 \hline
 \multicolumn{3}{|c|}{\textbf{Coefficients of} $T_2$} \\
 \hline
 $K_0$ & $ I_2(K_0)$  & $J_2(K_0)$ \\
 \hline
0.01 &$-0.19171684(4\pm 3)$   &  $-0.1917168(4\pm 1)$     \\ \hline
0.1  &-1.85828511(5$\pm$ 9)    &  -1.85828511(2$\pm$ 2)   \\ \hline
0.2&-3.3769476(8$\pm$ 1)     &  -3.3769476(8$\pm$ 4)      \\ \hline
0.3&-4.29785670058         &  -4.2978567(0 $\pm$ 6)       \\ \hline
0.4&-4.50418166(9 $\pm$ 8)   &   -4.5041816(6$\pm$ 2)        \\ \hline
0.5&-4.04299388345         &   -4.0429938(8$\pm$ 2)        \\ \hline
0.6&-3.1056196(0$\pm$ 2)     &  -3.1056196(0$\pm$ 2)         \\ \hline
0.7&-1.97241848(8$\pm$ 6)    &  -1.9724184(8$\pm$ 1)          \\ \hline
0.8&-0.9381858850(1$\pm$ 6)  &  -0.93818588(5$\pm$ 8)       \\ \hline
0.9&-0.23901332(5$\pm$ 3)    &  -0.23901332(5$\pm$ 9)       \\ \hline
0.99&-0.00233660398(4$\pm$ 4)&   -0.00233660(3 $\pm$ 7)   \\ \hline
\end{tabular}
 \caption{Numerical verification of the validity of equation \rf{eq1a}: the coefficients of $T_1$ and $T_2$ in \eqref{eq1alhs} and in  \rf{eq1arhs} agree up to the eighth order.}
\end{table}

\newpage

\subsection{Equation $V_{(2)}=U_{(2)}$}
\label{sec:tablesecondorder}
The tables below show the values of the coefficients of the algebra elements of (\ref{lhs2order}) and (\ref{rhs2order}). The fact that they agree implies on the validity of the equation obtained after expanding the Yang-Mills integral equation to second order in $\alpha$ and therefore, on the validity of the integral equation itself for any value of $\alpha$, at least up to that order.

\begin{table}[h]
\centering
\begin{tabular}{ |p{0.9cm}||p{4.5cm}|p{4.5cm}|  }
 \hline
 \multicolumn{3}{|c|}{\textbf{Coefficients of} $T_1$} \\
 \hline
 $K_0$& $I_1(K_0)+4\pi^2R_1(K_0)$ & $\mathcal{N}_{23}^{-}(K_0)$ \\ \hline
      0.01& 0.0106581(0$\pm$ 2)      & 0.010658107(5$\pm$ 3)\\ \hline
      0.1& 1.013915115(8$\pm$ 6)&1.01391511(5$\pm$ 4)\\ \hline
      0.2& 3.47985977(0$\pm$ 2)& 3.4798597(7$\pm$ 1)\\ \hline
      0.3& 6.0261611(3$\pm$ 8) &  6.0261611(4$\pm$ 6) \\ \hline
      0.4& 7.31167856(3$\pm$ 4)&  7.3116785(6$\pm$ 1)\\ \hline
      0.5& 6.7762177776(6$\pm$ 5)&6.77621777(7$\pm$ 9)\\ \hline
      0.6& 4.8526187817(7$\pm$ 6)&  4.85261878(1$\pm$ 4)\\ \hline
      0.7& 2.5695934(4$\pm$ 1) & 2.56959344117\\ \hline
      0.8& 0.8721002092(6$\pm$ 9)&  0.872100209(2$\pm$ 9)\\ \hline
      0.9& 0.115252473857&0.115252473(8$\pm$ 3)\\ \hline
      0.99& 0.000113925(3$\pm$ 3)&0.00011392533(9$\pm$ 1)\\ \hline
\end{tabular}
\begin{tabular}{ |p{0.9cm}||p{4.5cm}|p{4.5cm}|  }
 \hline
 \multicolumn{3}{|c|}{\textbf{Coefficients of} $T_2$} \\
 \hline
 $K_0$& $-(I_2(K_0)+4\pi^2R_2(K_0))$ & $\mathcal{N}_{13}^{-}(K_0)$ \\\hline
        0.01& -0.46994245(7$\pm$ 3) &-0.46994245(7$\pm$ 1) \\\hline
        0.1&-4.38319034(2$\pm$ 9) &-4.3831903(4$\pm$ 1)\\\hline
        0.2&-7.0567939(8$\pm$ 1) &-7.0567939(8$\pm$ 1) \\\hline
        0.3&-7.20852026286  &-7.2085202(6$\pm$ 1) \\\hline
        0.4&-5.26717871(7$\pm$ 8)&  -5.26717871(7$\pm$ 6)\\\hline
        0.5&-2.5185733549(6$\pm$ 9)& -2.51857335(4$\pm$ 2)\\\hline
        0.6&-0.2961954(3$\pm$ 2) &-0.29619543(4$\pm$ 7) \\\hline
        0.7& 0.71488621(2$\pm$ 6) &0.71488621(2$\pm$ 7)\\\hline
        0.8& 0.6675901030(5$\pm$ 6)&  0.66759010(3$\pm$ 4) \\\hline
        0.9&0.22170728(6$\pm$ 3)&0.22170728(6$\pm$ 1)\\\hline
        0.99& 0.00233492080(0$\pm$ 4)&0.00233492(0$\pm$ 8) \\ \hline
\end{tabular}
\caption{Comparison between the coefficients of $T_1$ and $T_2$ of equations (\ref{lhs2order}) and (\ref{rhs2order}).}
\end{table}
\newpage
\begin{table}[h]
\centering
\begin{tabular}{ |p{0.9cm}||p{4.5cm}|p{4.5cm}|  }
 \hline
 \multicolumn{3}{|c|}{\textbf{Coefficients of} $T_1^2$} \\
 \hline
 $K_0$& $4\pi S_1(K_0)$ & $\mathcal{N}_{1}(K_0)$ \\ \hline
          0.01& 78.894398(6$\pm$ 2)&78.8943986254 \\ \hline
          0.1&72.92922798(7$\pm$ 8)& 72.9292279882\\ \hline
          0.2&57.2526284(6$\pm$ 2) &57.2526284639 \\ \hline
          0.3&37.7395933527 &37.7395933527\\ \hline
          0.4&20.4073345(1$\pm$ 3) &20.4073345139\\ \hline
          0.5&8.690611129(8$\pm$ 2) & 8.69061112984\\ \hline
          0.6&2.7109923(7$\pm$ 3) &2.71099237533\\ \hline
          0.7&0.5387264(6$\pm$ 1) &0.538726461126\\ \hline
          0.8&0.0496472951(1$\pm$ 2) &0.0496472951165\\ \hline
          0.9&0.0007667310202(6$\pm$ 1) &0.000766731020264\\ \hline
          0.99&7.21318539905$\times 10^{-10}$   &7.213185398(9$\pm$ 2)$\times 10^{-10}$\\ \hline
\end{tabular}
\begin{tabular}{ |p{0.9cm}||p{4.5cm}|p{4.5cm}|  }
 \hline
 \multicolumn{3}{|c|}{\textbf{Coefficients of} $T_2^2$} \\
 \hline
 $K_0$& $4\pi S_2(K_0)$ & $\mathcal{N}_{2}(K_0)$ \\\hline
            0.01& 0.018377674(1$\pm$ 2)&0.0183776(7$\pm$ 1)\\
            0.1&1.7266117(8$\pm$ 1) &1.7266117(8$\pm$ 2) \\
            0.2&5.7018878(2$\pm$ 4) & 5.7018878(2$\pm$ 7) \\
            0.3&9.2357861093(6$\pm$ 1)&9.235786(1$\pm$ 1) \\
            0.4&10.1438262(5$\pm$ 4) & 10.1438262(5$\pm$ 5) \\
            0.5&8.172899770(8$\pm$ 4)&8.1728997(7$\pm$ 4) \\
            0.6&4.82243658089 &4.8224365(8$\pm$ 3) \\
            0.7&1.9452173(4$\pm$ 2)& 1.9452173(4$\pm$ 1)\\
            0.8&0.4400963774(2$\pm$ 5)&  0.4400963(7$\pm$ 1)\\
            0.9&0.0285636848378 & 0.0285636(8$\pm$ 1) \\
            0.99&2.72985908954e-06&2.729(8$\pm$ 8)e-06 \\\hline
\end{tabular}
\caption{Comparison between the coefficients of $T_1^2$ and $T_2^2$ of equations (\ref{lhs2order}) and (\ref{rhs2order}).}
\end{table}

\newpage

\begin{table}[h]
\centering
\begin{tabular}{ |p{0.9cm}||p{4.5cm}|p{4.5cm}|  }
 \hline
 \multicolumn{3}{|c|}{\textbf{Coefficients of} $\{T_1,T_2\}$} \\
 \hline
 $K_0$& $4\pi S_{12}(K_0)$ & $\mathcal{N}_{12}(K_0)$ \\ \hline
              0.01& -1.2041160882(4$\pm$ 2) & -1.20411608813 \\ \hline
              0.1&-11.2214288(0$\pm$ 5) &-11.2214288084\\ \hline
              0.2&-18.0678738(4$\pm$ 7) &-18.0678738427 \\ \hline
              0.3&-18.669622708 & -18.6696227077\\ \hline
              0.4&-14.3877884(2$\pm$ 1)&-14.3877884224\\ \hline
              0.5&-8.4277810668(8$\pm$ 6)&  -8.42778106684\\ \hline
              0.6&-3.6157418(0$\pm$ 4) & -3.6157418051 \\ \hline
              0.7&-1.02368943(3$\pm$ 4)&-1.02368943394 \\ \hline
              0.8&-0.1478160841(3$\pm$ 1)& -0.147816084139\\ \hline
              0.9&-0.0046798144427(1$\pm$ 3)& -0.00467981444135\\ \hline
              0.99&-4.43745194072$\times 10^{-8}$&  -4.437470(7$\pm$ 7)$\times 10^{-8}$\\ \hline
\end{tabular}
\caption{Comparison between the coefficients of $\{T_1,T_2\}$ of equations (\ref{lhs2order}) and (\ref{rhs2order}).}
\end{table}
\newpage
\begin{table}[h]
\centering
\begin{tabular}{ |p{0.9cm}||p{4.5cm}|p{4.5cm}|p{4.5cm}|  }
 \hline
 \multicolumn{4}{|c|}{\textbf{Coefficients of} $T_3$, $\{T_2,T_3\}$ and $\{T_1,T_3\}$} \\
 \hline
 $K_0$&  $4\pi^2R_3(K_0)+4\pi S_3(K_0)$& $\mathcal{N}_{23}^{+}(K_0)$ & $\mathcal{N}_{13}^{+}(K_0)$ \\ \hline
      0.01& $\pm$ 2 $\times$ $10^{-8}$&$\pm$ 7 $\times$ $10^{-11}$&$\pm$ 9 $\times$ $10^{-10}$\\\hline
      0.1&$\pm$ 7 $\times$ $10^{-8}$&$\pm$ 2 $\times$ $10^{-9}$&$\pm$ 6 $\times$ $10^{-9}$\\\hline
      0.2&$\pm$ 1 $\times$ $10^{-7}$&$\pm$ 7 $\times$ $10^{-9}$&$\pm$ 1 $\times$ $10^{-8}$\\\hline
      0.3&$\pm$ 2 $\times$ $10^{-11}$&$\pm$ 1 $\times$ $10^{-8}$&$\pm$ 4 $\times$ $10^{-9}$\\\hline
      0.4&$\pm$ 7 $\times$ $10^{-8}$&$\pm$ 1 $\times$ $10^{-8}$&$\pm$ 6 $\times$ $10^{-9}$\\\hline
      0.5&$\pm$ 7 $\times$ $10^{-10}$&$\pm$ 9 $\times$ $10^{-9}$&$\pm$ 7 $\times$ $10^{-9}$\\\hline
      0.6&$\pm$ 5 $\times$ $10^{-9}$&$\pm$ 4 $\times$ $10^{-9}$&$\pm$ 7 $\times$ $10^{-9}$\\\hline
      0.7&-4.44(5$\pm$ 2) $\times$ $10^{-11}$&$\pm$ 6 $\times$ $10^{-11}$&$\pm$ 5 $\times$ $10^{-9}$\\\hline
      0.8&-1$\pm$ 7$\times$ $10^{-11}$&$\pm$ 9 $\times$ $10^{-10}$&$\pm$ 3 $\times$ $10^{-9}$\\\hline
      0.9&-3.(7$\pm$ 7) $\times$ $10^{-13}$&$\pm$ 3 $\times$ $10^{-10}$&$\pm$ 5 $\times$ $10^{-9}$\\\hline
      0.99&9.599(0$\pm$ 4) $\times$ $10^{-16}$&-1.(6$\pm$ 2) $\times$ $10^{-11}$&$\pm$ 8$\times$ $10^{-9}$\\
       \hline
\end{tabular}
\caption{The coefficients above are the ones that should vanish in the equation obtained at second order in $\alpha$; indeed, within a numerical precision, they are zero.}
\end{table}


\newpage
\begin{thebibliography}{99}

\bibitem{ym1}
  L.~A.~Ferreira and G.~Luchini,
  ``Integral form of Yang-Mills equations and its gauge invariant conserved charges,''
  Phys.\ Rev.\ D {\bf 86}, 085039 (2012);
  [arXiv:1205.2088 [hep-th]].

\bibitem{ym2}
  L.~A.~Ferreira and G.~Luchini,
  ``Gauge and Integrable Theories in Loop Spaces,''
  Nucl.\ Phys.\ B {\bf 858}, 336 (2012);
  [arXiv:1109.2606 [hep-th]].

  \bibitem{thooft}
  G.~'t Hooft,
  ``Magnetic Monopoles in Unified Gauge Theories,''
  Nucl.\ Phys.\ B {\bf 79}, 276 (1974).
  doi:10.1016/0550-3213(74)90486-6

  \bibitem{polyakov}
  A.~M.~Polyakov,
  ``Particle Spectrum in the Quantum Field Theory,''
  JETP Lett.\  {\bf 20}, 194 (1974);
  [Pisma Zh.\ Eksp.\ Teor.\ Fiz.\  {\bf 20}, 430 (1974)].


  \bibitem{prasad}
  M.~K.~Prasad and C.~M.~Sommerfield,
   ``An Exact Classical Solution for the 't Hooft Monopole and the Julia-Zee Dyon,''
  Phys.\ Rev.\ Lett.\  {\bf 35}, 760-762 (1975).

\bibitem{bogo}
  E.~B.~Bogomolny,
  ``Stability of Classical Solutions,''
  Sov.\ J.\ Nucl.\ Phys.\  {\bf 24}, 449-454 (1976)
  [Yad.\ Fiz.\  {\bf 24}, 861 (1976)].
  
  \bibitem{manton}
  N.~S.~Manton and P.~Sutcliffe,
  ``Topological solitons,'' Cambridge Monographs on Mathematical Physics, 2004,
Cambridge University Press
  
  \bibitem{weinberg_book}
  E.~J.~Weinberg,
  ``Classical Solutions in Quantum Field Theory : Solitons and Instantons in High Energy Physics,''
  Cambridge Monographs on Mathematical Physics, 2012, Cambridge University Press


  \bibitem{us}
  C.~P.~Constantinidis, L.~A.~Ferreira and G.~Luchini,
  ``A remark on the asymptotic form of BPS multi-dyon solutions and their conserved charges,''
  {\em Journal of High Energy Physics}, JHEP {\bf 1512}, 137 (2015);
  doi:10.1007/JHEP12(2015)137;
  [arXiv:1508.03049 [hep-th]].
  
  \bibitem{ym_original}
  C.~N.~Yang and R.~L.~Mills,
  ``Conservation of Isotopic Spin and Isotopic Gauge Invariance,''
  Phys.\ Rev.\  {\bf 96} (1954) 191.
  doi:10.1103/PhysRev.96.191

  \bibitem{afs1}
  O.~Alvarez, L.~A.~Ferreira and J.~Sanchez Guillen,
  ``A New approach to integrable theories in any dimension,''
  Nucl.\ Phys.\ B {\bf 529}, 689 (1998);
  doi:10.1016/S0550-3213(98)00400-3;
  [hep-th/9710147].
  
  \bibitem{afs2}
  O.~Alvarez, L.~A.~Ferreira and J.~Sanchez-Guillen,
  ``Integrable theories and loop spaces: Fundamentals, applications and new developments,''
  Int.\ J.\ Mod.\ Phys.\ A {\bf 24}, 1825 (2009);
  doi:10.1142/S0217751X09043419;
  [arXiv:0901.1654 [hep-th]].
  




 




\end{thebibliography}
\end{document}